\documentclass{emulateapj}
\usepackage{apjfonts}
\usepackage{natbib}
\usepackage{lscape}

\def\chandra{{\it Chandra\/}}

\def\asca{{\it ASCA\/}}

\def\flux{ergs~cm$^{-2}$~s$^{-1}$}
\def\etal{{et\,al.\,}}
\begin{document}
\slugcomment{Data and images available at http://www.astro.psu.edu/users/niel/cdfs/cdfs-chandra.html}

\title{The {\bf {\em Chandra}} Deep Field-South Survey: 2 Ms Source Catalogs}

\author{
B.~Luo,\altaffilmark{1} 
F.~E.~Bauer,\altaffilmark{2} 
W.~N.~Brandt,\altaffilmark{1}
D.~M.~Alexander,\altaffilmark{3}
B.~D.~Lehmer,\altaffilmark{3}
D.~P.~Schneider,\altaffilmark{1}
M.~Brusa,\altaffilmark{4,5} 
A.~Comastri,\altaffilmark{6}
A.~C.~Fabian,\altaffilmark{7} 
A.~Finoguenov,\altaffilmark{4,5} 
R.~Gilli,\altaffilmark{6} 
G.~Hasinger,\altaffilmark{4}
A.~E.~Hornschemeier,\altaffilmark{8} 
A.~Koekemoer,\altaffilmark{9} 
V.~Mainieri,\altaffilmark{10} 
M.~Paolillo,\altaffilmark{11}
P.~Rosati,\altaffilmark{10} 
O.~Shemmer,\altaffilmark{1} 
J.~D.~Silverman,\altaffilmark{12} 
I.~Smail,\altaffilmark{13} 
A.~T.~Steffen,\altaffilmark{14} 
\& C.~Vignali\altaffilmark{15}
}
\altaffiltext{1}{Department of Astronomy \& Astrophysics, 525 Davey Lab, 
The Pennsylvania State University, University Park, PA 16802, USA}
\altaffiltext{2}{Columbia Astrophysics Laboratory, Columbia University, 
Pupin Laboratories, 550 W. 120th St., New York, NY 10027, USA}
\altaffiltext{3}{Department of Physics, Durham University, 
Durham, DH1 3LE, UK}
\altaffiltext{4}{Max-Planck-Institut f\"ur Extraterrestrische Physik, 
Giessenbachstrasse, D-85748 Garching b. M\"unchen, Germany}
\altaffiltext{5}{University of Maryland, Baltimore County, 1000 
Hilltop Circle, Baltimore, MD 21250, USA}
\altaffiltext{6}{INAF---Osservatorio Astronomico di Bologna, Via Ranzani 1, 
Bologna, Italy}
\altaffiltext{7}{Institute of Astronomy, Madingley Road, Cambridge, 
CB3 0HA, UK}
\altaffiltext{8}{Laboratory for X-ray Astrophysics, NASA Goddard Space 
Flight Center, Code 662, Greenbelt, MD 20771, USA}
\altaffiltext{9}{Space Telescope Science Institute, 3700 San Martin Drive, 
Baltimore, MD 21218, USA}
\altaffiltext{10}{European Southern Observatory, Karl-Schwarzschild-Strasse 2, 
Garching, D-85748, Germany}
\altaffiltext{11}{Dipartimento de Scienze Fisiche, Universit\`a di Napoli, 
Via Cinthia, 80126 Napoli, Italy}
\altaffiltext{12}{ETH Zurich, Institute of Astronomy, Department of Physics, 
Wolfgang-Pauli-Strasse 16, 8093 Zurich, Switzerland}
\altaffiltext{13}{Institute of Computational Cosmology, Durham University, 
Durham, DH1 3LE, UK}
\altaffiltext{14}{Spitzer Science Center, California Institute of Technology, 
Mail Code 220-6, 1200 East California Boulevard, Pasadena, CA 91125, USA}
\altaffiltext{15}{Universit\'a di Bologna, Via Ranzani 1, Bologna, Italy}

\begin{abstract}
We present point-source catalogs for the $\approx2$~Ms exposure of 
the \chandra\ Deep Field-South (\hbox{CDF-S}); this is one of the 
two most-sensitive \hbox{X-ray} surveys ever performed. The survey covers an
area of $\approx436$ arcmin$^{2}$ and reaches on-axis sensitivity limits of 
$\approx1.9\times10^{-17}$ and $\approx1.3\times10^{-16}$ \flux\ 
for the \hbox{0.5--2.0} and 2--8~keV bands, respectively. Four hundred and sixty-two
\hbox{X-ray} point sources are detected in at least one of 
three \hbox{X-ray} bands that were searched; 
135 of these sources are new compared to the previous $\approx1$~Ms
\hbox{CDF-S} detections. Source positions are determined using centroid
and matched-filter techniques; the median positional uncertainty is 
$\approx0\farcs36$. The X-ray--to--optical flux ratios of the newly detected
sources indicate a variety of source types; $\approx$55\% of them appear to
be active galactic nuclei while $\approx$45\% appear to be starburst 
and normal galaxies. In addition to the main
\chandra\ catalog, we provide a supplementary catalog of 86 \hbox{X-ray}
sources in the $\approx2$~Ms \hbox{CDF-S} footprint that was created 
by merging the $\approx250$~ks 
Extended \chandra\ Deep Field-South with the 
\hbox{CDF-S}; this approach provides 
additional sensitivity in the outer portions
of the \hbox{CDF-S}. A second supplementary catalog that contains 
30 \hbox{X-ray} 
sources was constructed by matching lower significance \hbox{X-ray} sources
to bright optical counterparts ($R<23.8$); the majority of these sources
appear to be starburst and normal galaxies. 
The total number of sources in the main and supplementary catalogs is 578.
$R$-band optical counterparts and basic optical and infrared photometry 
are provided for the \hbox{X-ray} sources in the main and 
supplementary catalogs. 
We also include existing spectroscopic redshifts for 224 of the X-ray sources.
The average backgrounds in the
0.5--2.0 and 2--8~keV bands are 0.066 and 0.167~counts~Ms$^{-1}$~pixel$^{-1}$,
respectively, and the background counts follow Poisson distributions.
The effective exposure times and sensitivity limits of 
the \hbox{CDF-S} are now comparable 
to those of the $\approx2$~Ms \chandra\ Deep Field-North (\hbox{CDF-N}). 
We also present
cumulative number counts for the main catalog and compare the results 
to those for
the \hbox{CDF-N}.
The soft-band number counts for these two fields 
agree well with each other at fluxes higher than 
$\approx2\times10^{-16}$~\flux, 
while the \hbox{CDF-S}
number counts are up to $\approx25\%$ smaller
than those for the \hbox{CDF-N}
at fluxes below $\approx2\times10^{-16}$~\flux\ in the soft
band and $\approx2\times10^{-15}$~\flux\ in the hard band,
suggesting small field-to-field variations. 
\end{abstract}

\section{INTRODUCTION}

One of the greatest successes of the {\it Chandra X-Ray Observatory} 
(\chandra) has been the characterization of the
sources creating the 0.5--8 keV cosmic \hbox{X-ray} background (CXRB), and the deepest
\chandra\ surveys form a central part of this effort. The two 
deepest \chandra\ surveys, the \chandra\ Deep Field-North and \chandra\ 
Deep Field-South (\hbox{CDF-N} and \hbox{CDF-S}, jointly CDFs; see 
\citealt{Brandt2005} for a review), have each detected hundreds of 
\hbox{X-ray} sources over $\approx 450$~arcmin$^2$ areas with enormous 
multiwavelength observational investments. They have measured the highest 
sky density of accreting supermassive black holes (SMBHs) to date and have 
also enabled novel X-ray studies of starburst and normal galaxies, groups 
and clusters of galaxies, large-scale structures in the distant universe, 
and Galactic stars. 

As part of an effort to create still deeper \hbox{X-ray} surveys, we proposed
for substantial additional exposure on the \hbox{CDF-S} during \chandra\ 
Cycle~9. The \hbox{CDF-S} has superb and improving coverage at optical, 
infrared, and radio wavelengths; it will continue to be a premiere 
multiwavelength deep-survey field for the coming decades as additional 
large facilities are deployed in the southern hemisphere. Furthermore, 
owing to the 1~Ms of \chandra\ exposure already available 
\citep[][hereafter G02]{Giacconi2002}, the 
\hbox{CDF-S} is a natural field to observe more sensitively. 
Although our 
proposal was not approved in the peer review, subsequently 
1~Ms of Director's Discretionary Time was allocated 
for deeper \hbox{CDF-S} 
observations. The allocated observations were successfully executed in 
2007 September, October and November, raising the \hbox{CDF-S} 
exposure to $\approx 2$~Ms and 
improving its sensitivity to be comparable to that of the \hbox{CDF-N} 
\citep[e.g.,][hereafter A03]{Alexander2003}.
Additional sky coverage at such flux levels 
is critically important as it substantially improves the statistical sample 
sizes of the faintest X-ray sources and also allows a basic assessment of 
the effects of cosmic variance. Furthermore, approximately doubling the 
exposure on previously detected sources substantially improves the constraints
on their positions, spectral properties, and variability properties.

In this paper, we present up-to-date \chandra\ source catalogs and data 
products derived from the full $\approx 2$~Ms \hbox{CDF-S} data set along 
with details of the observations, data processing, and technical analysis. 
Detailed subsequent investigations and scientific
interpretation of the new \hbox{CDF-S} sources will be presented in future
papers, e.g., studies of heavily obscured and Compton-thick 
active galactic nuclei (AGNs), high-redshift AGNs, AGN spectra and variability, 
starburst and normal galaxies, and clusters and groups of galaxies.
In \S2 we describe the observations and data reduction, and in \S3 we
present the main and supplementary point source catalogs and 
describe the methods used to create these catalogs. In \S4 we estimate
the background and sensitivity across the survey region. We also present 
basic number-count results for point sources in \S5. We summarize in \S6.

The Galactic column density along the line of sight to the 
\hbox{CDF-S} is remarkably low: 
\hbox{$N_{\rm H}=8.8\times 10^{19}$~cm$^{-2}$} \citep[e.g.,][]{Stark1992}.
The coordinates throughout this paper are J2000. 
A $H_0=70$~km~s$^{-1}$~Mpc$^{-1}$, $\Omega_{\rm M}=0.3$, and 
$\Omega_{\Lambda}=0.7$ cosmology is adopted.

\section{OBSERVATIONS AND DATA REDUCTION}

\subsection{Observations and Observing Conditions}
The \hbox{CDF-S} consists of 23 separate observations described
in Table~\ref{tbl-obs}. The $\approx 1$~Ms catalogs for the first 11 
observations taken between 1999 October 14 and 2000 December 23 
were presented in G02 and A03. 
Note that observation 581 (1999 October 14)
was excluded from the data reduction and is not listed in Table~\ref{tbl-obs}
due to telemetry saturation and other problems. The second $\approx 1$~Ms 
exposure consisted of 12 observations taken between 2007 September 20 and
2007 November 4.

The Advanced CCD Imaging Spectrometer imaging array (ACIS-I; 
\citealt{Garmire2003}) was used
for all of the \chandra\ observations.
The ACIS-I is composed of 
four $1024\times1024$ pixel CCDs (CCDs \hbox{I0--I3}), covering a 
field of view of $16\farcm 9\times 16\farcm 9$ ($\approx$285
arcmin$^2$), and the pixel size of the CCDs is $\approx$$0\farcs 492$. 
The focal-plane 
temperature was $-110\degr$C for observations 1431-0 and 1431-1, and 
$-120\degr$C for the others. The 12 new observations were taken in Very
Faint mode to improve the screening of background events and thus increase
the sensitivity of ACIS in detecting faint \hbox{X-ray} sources 
\citep{Vikhlinin2001}.

The background light curves for all 23 observations were inspected using
EVENT BROWSER in the Tools for ACIS Real-time Analysis ({\sc TARA}; 
\citealt{Broos2000}) software package. Aside from a mild flare 
during observation 1431-0
(factor of $\approx3$ increase for $\approx5$~ks), all data sets are free 
from significant
flaring, and the background is stable within $\approx$20\% of 
typical quiescent {\it Chandra}
values. After filtering on good-time intervals and removing the one mild
flare, we are left with 1.911~Ms of total exposure time for the 
23 observations.

Because of the differences in pointings and roll angles for the individual
exposures, the total region
covered by the entire \hbox{CDF-S} is 435.6 arcmin${^2}$, considerably
larger than the ACIS-I field of view. Combining the 23 observations, the 
average aim point (weighted by exposure time) is $\alpha_{\rm J2000.0}=
03^{\rm h}32^{\rm m}28\fs80$, $\delta_{\rm J2000.0}=-27\degr48\arcmin23\farcs0$.

\subsection{Data Reduction}

The basic archive data products were processed with the
\chandra\ X-ray Center (CXC) pipeline software versions listed in Table 1.
The reduction and analysis of the data used \chandra\ Interactive Analysis
of Observations (CIAO) tools whenever possible\footnote{See
http://cxc.harvard.edu/ciao/ for details on CIAO.}; however, custom
software, including the {\sc TARA} package, was also used. Each
observation was
reprocessed using the CIAO tool {\sc acis\_process\_events}, to correct for the
radiation damage sustained by the CCDs during the first few months of
\chandra\ operations using a Charge Transfer Inefficiency (CTI)
correction procedure \citep{Townsley2000,Townsley2002}
\footnote{Note that the CXC CTI correction procedure
is only available for $-120\degr$C data; thus we did not CTI-correct
observations 1431-0 and 1431-1.}, to remove the standard pixel
randomization which blurs the {\it Chandra} point spread function (PSF),
and to apply a modified bad-pixel file as detailed below.

One important deviation from the standard {\it Chandra} reduction
procedure outlined by the CXC is implementation of a stripped-down 
bad-pixel file. 
We note that the standard bad-pixel file supplied with all
{\it Chandra} data currently excludes $\approx$ 6--7\% of the total
effective area on front-illuminated devices (e.g., ACIS-I). A large
fraction of the bad-pixel locations identified in this file, however,
appear to be flagged solely because they show a few extra events (per Ms)
almost exclusively below 0.5--0.7 keV.\footnote{See
http://cxc.harvard.edu/cal/Acis/Cal\_prods/badpix/index.html} Good events
with energies above 0.7~keV that fall on these bad pixels are likely to be
perfectly acceptable for source searching, as well as for photometry and
spectral analysis albeit with a few mild caveats regarding
misinterpretation. Rather than reject all events falling on such columns,
we instead adopted a procedure to only exclude events below a
row-dependent energy of 0.5--0.7 keV.
\footnote{The energy range of 0.5--0.7~keV
and frequency of occurrence were verified by visual inspection of such
columns in our $\approx2$~Ms data set. We found that such ``hot'' soft 
columns were
not clearly seen in any individual observations. The upper energy bound
appears to vary as a function of distance from the readout edge of the
front-illuminated CCDs, such that rows closest to the readout edge only
have extra events below $\approx$0.5~keV, while those furthest away have
extra events extending up to $\approx$0.7~keV.} To this end, we generated
a stripped-down bad-pixel file, only selecting obvious bad columns and pixels
above 1~keV; this excluded $\approx1.5\%$ of the total effective area on
front-illuminated devices. Once the entire $\approx2$~Ms data set 
was combined, we
isolated ``hot'' soft columns as those where the total number of events with
energies below 0.7~keV was 5$\sigma$ or more above the mean. 
We then rejected
any events in those columns that fell below a row-dependent
0.5--0.7~keV; this removed 1\% of all events.

Through inspection of the data in CCD coordinates, we
additionally discovered that the CXC-preferred CIAO tool {\sc 
acis\_run\_hotpix}
failed to flag a substantial number of obvious cosmic-ray afterglows
($\sim$100--200 per observation, depending on exposure length),
elevating the overall background and, in egregious cases, leaving
afterglows to be mistaken as real sources. 
This problem appeared to be worse for
Faint mode data, presumably because the additional $5\times5$ screening
applied in Very Faint mode rejects the strongest afterglows 
\citep{Vikhlinin2001}.
To remedy this situation, we reverted to using the more stringent
{\sc acis\_detect\_afterglow} algorithm on all of our data. 
Notably, none of our
sources has a count rate high enough that {\sc acis\_detect\_afterglow} would
reject true source counts, which we verified by inspection of events
flagged by this routine. Even {\sc acis\_detect\_afterglow} failed
to reject all afterglows, and thus we created custom
software to remove many remaining faint afterglows from
the data. Working in CCD coordinates, we removed additional faint
afterglows with three or more total counts occuring within 20~s (or
equivalently 6 consecutive frames). In total, we removed 229 
total events associated with afterglows.
In all cases, we inspected the data set and found that such flagged events were
isolated and not associated with apparent legitimate X-ray sources.

\section{PRODUCTION OF THE POINT-SOURCE CATALOGS}
The production of the point-source catalogs largely followed the procedure
described in \S3 of A03. The main differences in the catalog-production 
procedure used here are the following:

\begin{enumerate}

\item
Our main \chandra\ catalog includes sources detected by running {\sc
wavdetect} \citep{Freeman2002} at a false-positive probability 
threshold of 10$^{-6}$, less
conservative than the 10$^{-7}$ value adopted by A03. Even with this
revised threshold, we expect the fraction of false sources to be small; 
see \S3.2 for details.

\item
Additional sensitivity can be obtained by merging the $\approx 250$~ks
Extended \chandra\ Deep Field-South (\hbox{E-CDF-S}; \citealt{Lehmer2005}, 
hereafter L05) with the $\approx2$~Ms \hbox{CDF-S}. An additional 86
\hbox{X-ray} sources were detected with this approach. 
These sources are presented
in a supplementary catalog described in \S3.3.2.

\end{enumerate}

\subsection{Image and Exposure Map Creation}
We registered the observations in the following manner. {\sc wavdetect}
was run on each individual cleaned image to generate an initial source list.
Centroid positions for each detected source were determined using the
reduction tool
{\sc acis extract} (AE; \citealt{Broos2000}).\footnote{The {\sc acis extract} 
software can be accessed from
http://www.astro.psu.edu/xray/docs/TARA/ae\_users\_guide.html}
The observations were registered to a common astrometric
frame by matching X-ray centroid positions to optical sources detected in
deep $R$-band images taken with the Wide Field Imager (WFI) of the MPG/
ESO telescope at La Silla (see \S2 of \citealt{Giavalisco2004}). The
matching was performed using the CIAO tools {\sc reproject\_aspect} and
{\sc wcs\_update} adopting a 3$\arcsec$ matching radius and a 
residual rejection limit \footnote{This is a parameter used in {\sc wcs\_update}
to remove source pairs based on pair positional offsets.}
of 0\farcs6; 50--100 sources were typically used in each observation 
for the final
astrometric solution. The tool {\sc wcs\_update} applied linear translations
ranging from 0\farcs05 to 0\farcs34, rotations ranging from $-0\fdg239$ to
0\fdg009, and scale stretches ranging from 0.999563 to 1.000714; individual
registrations are accurate to $\approx$0\farcs3. All of the observations were
then reprojected to the frame of observation 2406, since this data set required
the smallest translation to align it with the optical astrometric frame.

\begin{figure}
\centerline{
\includegraphics[scale=0.4]{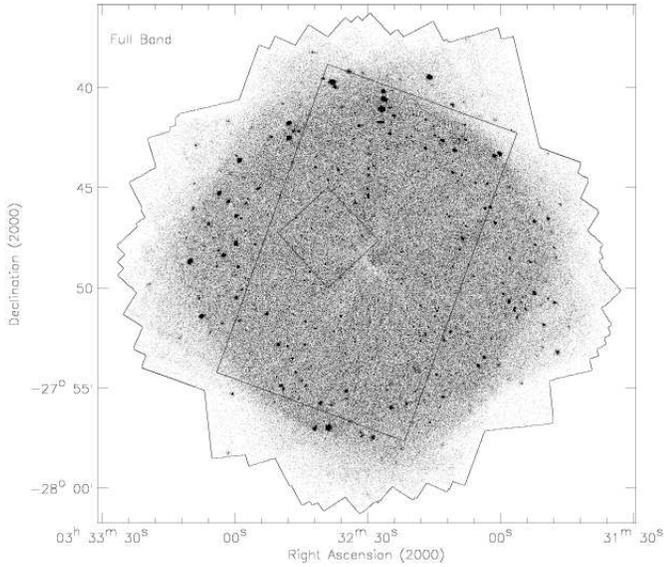}
}
\figcaption{
Full-band (0.5--8.0~keV) raw image of the $\approx$2~Ms \hbox{CDF-S}.
The gray scales are linear.
The apparent scarcity of sources near the field center is largely due
to the small PSF at that location 
(see Figs.~\ref{clrimg} and \ref{pos} for clarification).
The black outline surrounding the image indicates
the extent of all the \hbox{CDF-S} observations.
The large rectangle indicates the GOODS-S
\citep{Giavalisco2004} region, and
the central square indicates the {\it Hubble} Ultra Deep Field
(UDF; \citealt{Beckwith2006}) region. The cross near the center of the images
indicates the average aim point, weighted by exposure time (see
Table~\ref{tbl-obs}).\label{fbimg}}
\end{figure}

\begin{figure}
\centerline{
\includegraphics[scale=0.4]{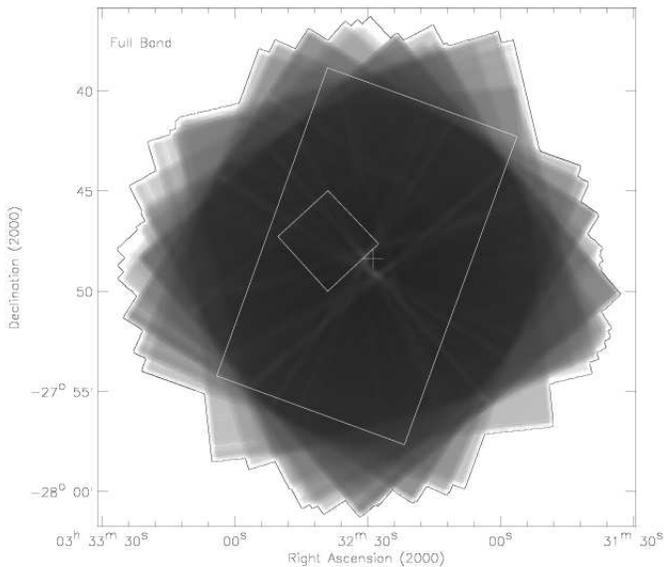}
}
\figcaption{
Full-band (0.5--8.0~keV) exposure map of 
the $\approx2$~Ms \hbox{CDF-S}.
The darkest areas represent the highest effective 
exposure times (the maximum value is 1.884~Ms). The gray scales
are logarithmic.
The regions and the cross symbol have the same meaning as those in
Fig. \ref{fbimg}.
\label{fbemap}}
\end{figure}

\begin{figure}
\centerline{
\includegraphics[scale=0.4]{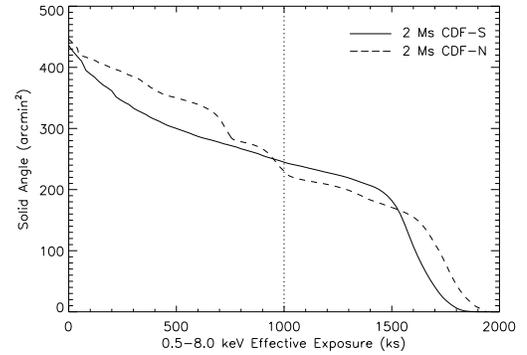}
}
\figcaption{
Amount of survey solid angle having at least a given amount of
full-band effective exposure for the $\approx2$~Ms \hbox{CDF-S}
({\it solid curve}).
The maximum exposure is $\approx1.884$~Ms.
The vertical dotted
line shows an effective exposure of 1~Ms. About 245 arcmin$^2$ 
($\approx 56\%$) of the \hbox{CDF-S} survey area has $>1$~Ms 
effective exposure. Corresponding data from 
the $\approx2$~Ms \hbox{CDF-N} (A03) are
plotted as a dashed curve for comparison.
\label{emapcum}}
\end{figure}

We constructed images using the standard
\asca\ grade set (\asca\ grades 0, 2, 3, 4, 6) for three standard bands: 
\hbox{0.5--8.0~keV} (full band; FB), \hbox{0.5--2.0~keV}
(soft band; SB), and \hbox{2--8~keV} (hard band; HB). 
Figure~\ref{fbimg} shows the full-band raw image.
Exposure maps in the
three standard bands were created following the basic procedure
outlined in \S3.2 of \citet{Hornschemeier2001} and were normalized to the
effective exposures of a source located at the average aim point. 
Briefly, this
procedure takes into account the effects of vignetting, gaps between the CCDs,
bad-column filtering, bad-pixel filtering, and the spatially 
dependent degradation in quantum efficiency
due to contamination on the ACIS optical-blocking filters.
A photon index of
$\Gamma=1.4$ was assumed in creating the exposure maps, which is
approximately
the slope of the \hbox{X-ray} background in the
\hbox{0.5--8.0~keV} band 
\citep[e.g.,][]{Marshall1980,Gendreau1995,Hasinger1998}.
We show the full-band exposure map in Figure~\ref{fbemap}.
Using the full-band exposure map, we calculated the 
survey solid angle as a 
function of the minimum full-band effective exposure; the result is
plotted in Figure~\ref{emapcum}. Approximately 56\% and 42\% of the 
\hbox{CDF-S} field has a full-band effective exposure 
greater than 1~Ms and 1.5~Ms, 
respectively, with a maximum effective exposure of $\approx1.884$~Ms
(note this is slightly smaller than the 1.911~Ms total exposure since the
aim points of all the \chandra\ observations were not exactly the same).
The survey solid angles are comparable to those of the 
$\approx2$~Ms \hbox{CDF-N} (A03; dashed curve in Fig.~\ref{emapcum}).

\begin{figure}
\centerline{
\includegraphics[scale=0.4]{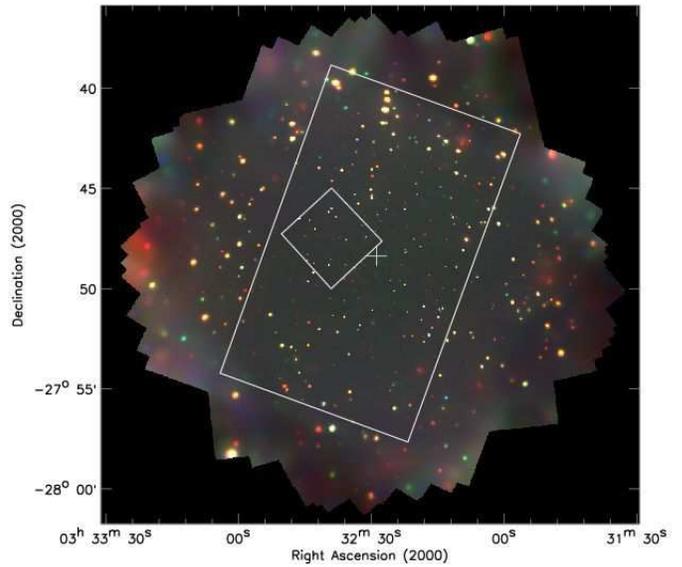}
}
\figcaption{
\chandra\ ``false-color'' image of the $\approx$2~Ms \hbox{CDF-S}. This image
is a color composite of the  exposure-corrected adaptively smoothed images
in the 0.5--2.0 keV ({\it red}), 2--4 keV ({\it green}), and 4--8
keV ({\it blue}) bands. 
The apparent smaller size and lower brightness of sources near the 
field center is due to the small PSF at that location. 
The regions and the cross symbol have the same meaning as those in
Fig. \ref{fbimg}.
\label{clrimg}}
\end{figure}

Adaptively smoothed images were created using the CIAO
tool {\sc csmooth} on the raw images. 
Exposure-corrected smoothed images were then constructed following
\S3.3 of \citet{Baganoff2003}.
We show in 
Figure~\ref{clrimg} a color composite of the exposure-corrected 
smoothed images
in the 0.5--2.0 keV ({\it red}), 2--4 keV ({\it green}), and 4--8 
keV ({\it blue}) bands.
Source searching was performed using only the raw images, while
many of the detected \hbox{X-ray} sources are shown
more clearly in the adaptively smoothed images.



\subsection{Point-Source Detection}

Point-source detection was performed in each of the three standard bands
with {\sc wavdetect} using a
``$\sqrt{2}$~sequence'' of wavelet scales (i.e.,\ 1, $\sqrt{2}$, 2,
$2\sqrt{2}$, 4, $4\sqrt{2}$, 8, $8\sqrt{2}$, and 16 pixels). 
The criterion for source
detection is that a source must
be found with a given false-positive probability threshold in at least one of
the three standard bands. For the main \chandra\ source catalog discussed in
\S3.3.1, the false-positive probability threshold in
each band was set to $1\times 10^{-6}$.

If we conservatively consider the three images searched 
to be independent, $\approx$18 false detections are expected in the main
\chandra\ source catalog for the case of a uniform background.
However, this false-source estimate is conservative, since 
a single pixel usually should not
be considered a source-detection cell, particularly at large off-axis angles
({\sc wavdetect} suppresses fluctuations on scales smaller than the PSF).
As quantified in
\S3.4.1 of A03, the number of false-sources is likely
\hbox{$\approx$2--3} times less than our conservative estimate.
We also provide additional source-significance information 
by running {\sc wavdetect} using false-positive probability
thresholds of $1\times 10^{-7}$ and $1\times 10^{-8}$. These results
are presented in $\S$3.3.1, which can be utilized
to perform more conservative source screening if desired.

\subsection{Point-Source Catalogs}

\subsubsection{Main Chandra Source Catalog}

The source 
lists resulting from the {\sc wavdetect} runs discussed in \S3.2 
with false-positive probability 
threshold of $1\times 10^{-6}$ were 
merged to create the main point-source catalog presented in 
Table~\ref{tbl-mcat}, which consists of 462 point sources. 
Whenever possible, we have quoted 
the position determined in 
the full band; when a source is not detected in the full band, we used, in 
order of priority, the soft-band position or hard-band position. 
For cross-band 
matching, we used a matching radius of $2\farcs 5$ for sources 
within $6\arcmin$ of the average aim point and $4\farcs 0$ for 
larger off-axis angles. These matching radii were chosen by
inspecting histograms showing the number of matches obtained as 
a function of angular separation \citep[e.g., see \S2 of][]{Boller1998}; 
the mismatch probability is $\la 1\%$ over the entire field. 
A few mismatches near the edge of the field were removed through 
visual inspections. 

We improved the {\sc wavdetect} source positions using the centroid and
matched-filter positions generated with AE. 
The centroid is simply the mean position of all
events within the AE extraction region, while the matched-filter position
is the
position found by correlating the full-band image in the vicinity of each
source with a combined PSF. The combined PSF is produced by combining the
``library'' PSF of a
source for each observation, weighted by the number of detected
counts.\footnote{The PSFs are
taken from the CXC PSF library; see
http://cxc.harvard.edu/ciao/dictionary/psflib.html.} This technique takes
into account the fact that, due to the complex PSF at large off-axis
angles, the X-ray source position is not always located at the peak of the
X-ray emission. The {\sc wavdetect}, centroid, and matched-filter techniques
provide comparable accuracy on-axis, while the matched-filter technique 
performs better off-axis. We chose the matched-filter positions as
our default, and then visually inspected each source. 
When the adopted position appeared to deviate from the
apparent center of the source by more than 0\farcs1, 
we modified the position manually such that it was
visually consistent with the apparent center.

We refined the absolute \hbox{X-ray} source positions
by matching the \hbox{X-ray} sources in the main \chandra\ catalog
to the WFI $R$-band optical sources (see \S3.1). 
There are $\approx30\,000$ optical sources across the \hbox{CDF-S} field,
which have accurate positions with positional error 
$\Delta_{\rm o}\approx0\farcs1$.\footnote{See 
http://archive.stsci.edu/pub/hlsp/goods/v1/h\_goods\_v1.0\_rdm.html.}
We selected relatively bright
optical sources with AB magnitudes $R\le 24$ ($\approx5\,500$ sources), 
and matched them to the 
\hbox{X-ray} sources using a 2$\farcs$5
matching radius. 
There are eight cases where one \hbox{X-ray} source has
two optical counterparts. The $R$-band 
magnitudes of the two counterparts differ by less than one in all cases, and
thus we selected the closer one 
as the most-probable counterpart. 
We also visually inspected the optical counterparts and, 
for purposes of positional checking, only keep those
sources that are point-like or slightly extended; ten extended sources 
were removed. Under these criteria, 
229 \hbox{X-ray} sources have bright optical counterparts.
We estimated the expected number of false matches by
manually shifting the \hbox{X-ray} source coordinates in right 
ascension and declination by $5\farcs0$ (both positive and negative
shifts) and recorrelating with the optical sources. On average, the 
number of false matches is $\approx35$ ($\approx15\%$), and the median
offset of these false matches is $\approx$1$\farcs$71. By comparing
the \hbox{X-ray} and optical source positions,
we found small shift and plate-scale corrections. These corrections have
been applied to the positions of all the \hbox{X-ray} sources
in the main and supplementary catalogs, resulting in
small ($<0\farcs2$) astrometric shifts.

\begin{figure}
\centerline{
\includegraphics[scale=0.5]{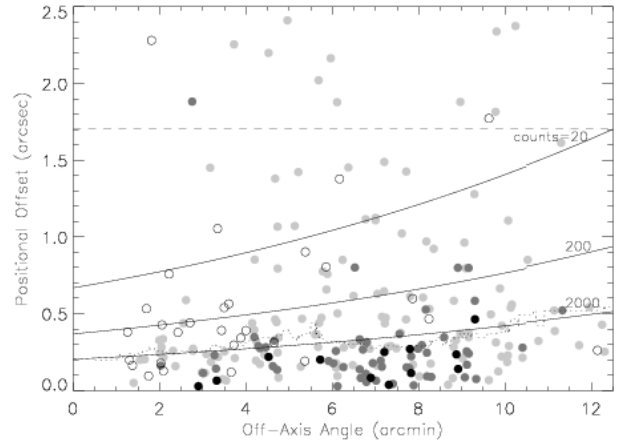}
}
\figcaption{
Positional offset vs. off-axis angle for
sources in the main {\it Chandra} catalog that were matched to WFI $R$-band
optical sources with AB magnitude $R\le24$ to within 2$\farcs$5. 
Black, dark gray, light gray, and open circles represent {\it
Chandra} sources with $\ge2000$, $\ge200$, $\ge20$, and $<20$ 
counts in the energy
band where the source position was determined, respectively.
The dotted curve shows the running median of all sources in
bins of 2\arcmin. The median offset of the expected false matches 
($\approx$1$\farcs$71) is indicated by the 
dashed line. These data were
used to derive the $\approx85\%$ confidence-level positional uncertainties 
of the \hbox{X-ray} sources in the main catalog; see eq. (1).
Three solid curves indicate the $\approx85\%$ confidence-level positional 
uncertainties for sources with counts of 20, 200 and 2000.
The number of black, dark-gray, and light-gray circles lying below/above
their corresponding solid curves are 11/1, 48/4 and 116/20, respectively.
Note that sources with more than 20 or 200 counts will have expected 
positional 
uncertainties smaller than those indicated by the corresponding solid curves.
\label{dpos}}
\end{figure}

\begin{figure}
\centerline{
\includegraphics[scale=0.5]{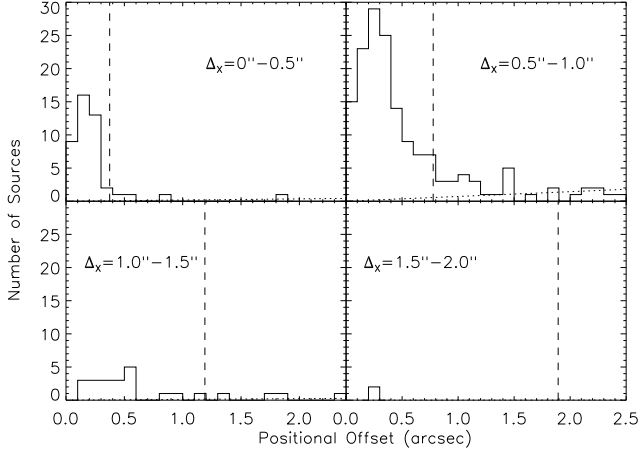}
}
\figcaption{
Histograms showing the distributions of positional offset for
sources in the main {\it Chandra} catalog that were matched to WFI $R$-band 
optical sources with $R\le24$ to within 2$\farcs$5. X-ray sources were
divided into four bins based on their positional uncertainties estimated
using eq. (1): $0\arcsec$--$0\farcs5$, $0\farcs5$--$1\farcs0$,
$1\farcs0$--$1\farcs5$, and $1\farcs5$--$2\farcs0$. 
The vertical dashed line indicates the median positional uncertainty for
X-ray sources in each bin.
Dotted lines show how many random $R\le24$ optical sources
are expected as a function of the positional offset.
Less than 20\% of
the optical counterparts lie beyond the median X-ray positional uncertainties
in all cases. 
\label{poshist}}
\end{figure}

We investigated the accuracy of the \hbox{X-ray} source positions
using these 229 \hbox{X-ray} detected bright optical sources.
Figure~\ref{dpos} shows the positional offset between 
the \hbox{X-ray} sources and their optical
counterparts as a function of the off-axis angle. The median offset
is $\approx$0$\farcs$36. However,
there are clear off-axis angle and source-count
dependencies. The off-axis angle dependence is due to the degradation of
the \chandra\ PSF
at large off-axis angles, while the count dependence is due to the
difficulty of finding the centroid of a faint \hbox{X-ray} source. 
Simulations have shown that the offsets of {\sc wavdetect} positions appear
to increase exponentially with off-axis angle and decrease with the 
number of source
counts in a power-law form \citep[e.g.,][]{Kim2007}.
Based on Figure~\ref{dpos} and taking into account the 
probability of false matches, we derived an empirical relation 
for the positional uncertainties
of the \hbox{X-ray} sources in our sample, which is 
\begin{equation}
\log \Delta_{\rm X}=0.0326 \theta-0.2595\log C+0.1625~,
\end{equation}

\noindent
where $\Delta_{\rm X}$ is the positional uncertainty in arcseconds, $\theta$ 
the off-axis angle in arcminutes, and $C$ the source counts in the energy
band where the source position was determined. 
We set an upper limit of 2000 on $C$ as the positional accuracy
does not improve significantly beyond that level.
Positional uncertainties
for $C=20$, 200, and 2000 are shown in Figure~\ref{dpos}.
The stated positional uncertainties are for the 
$\approx85\%$ confidence level, and are smaller than the {\sc wavdetect}
positional errors, especially at large off-axis angles, 
because of our positional refinement described above.
A few sources in Figure~\ref{dpos} have unexpectedly
large positional offsets; they could be false matches.\footnote{For example, 
the source with $>200$ counts and a positional offset of $\approx1\farcs9$ in 
Figure~\ref{dpos} is source
``289'' in the main \chandra\ catalog (see Table~2). This source does not have 
any optical counterpart after adopting a more appropriate matching radius,
as shown in the catalog.} There is also the possibility that a few of
them are off-nuclear \hbox{X-ray} sources 
\citep[e.g.,][]{Hornschemeier2004,Lehmer2006}.
Figure~\ref{poshist} shows the distributions of the positional 
offsets in four bins of different X-ray positional uncertainties, as well
as the expected numbers of false matches assuming a uniform spatial 
distribution of the $R\le 24$ optical sources. These histograms illustrate
clearly the reliability of our positional error estimates calculated using 
equation~(1).


The main {\it Chandra} \hbox{X-ray} source catalog 
is presented in Table~2, 
with the details of the columns given below.

\begin{enumerate}

\item
Column~1: the source number. Sources are listed in order of
increasing right ascension.

\item Columns~2 and 3: the right ascension and declination of 
the \hbox{X-ray} source,
respectively. These positions have been determined following the procedure 
described above. To avoid truncation error, we
quote the positions to higher precision than in the International Astronomical
Union (IAU) registered names beginning with the acronym ``CXO CDFS''. 

\item Column~4: the $\approx85\%$ confidence-level 
positional uncertainty in arcseconds. 
As shown above, the positional
uncertainty depends on off-axis angle and the number of detected counts,
and is estimated following equation (1). The minimum positional uncertainty
is $\approx0\farcs23$ for sources in the main catalog, and the 
maximum value is $\approx1\farcs90$.

\item Column~5: the off-axis angle of
the \hbox{X-ray} source in arcminutes. This is
calculated using the source position given in columns~2 and 3 and the 
average aim point of the \hbox{CDF-S} (see Table~1).

\item Columns~6--14: the source counts and the corresponding $1\sigma$
statistical errors \citep{Gehrels1986} or the upper limits on source counts
for the three standard bands, respectively.
The entries have not been corrected for vignetting. Source counts and 
statistical
errors have been calculated using circular-aperture photometry; extensive
testing has shown that this method is more reliable than the {\sc wavdetect}
photometry (e.g., \citealt{Brandt2001}; A03). 
The circular aperture was centered at the position given in
columns 2 and 3 for all bands. 
We have also computed photometry using AE, and the results are
in good agreement with this circular-aperture photometry.

The local background is determined in an annulus outside of the
source-extraction region. The mean number of background counts per pixel is
calculated from a Poisson model using ${n_1}/{n_0}$, where $n_0$ is the
number of pixels with 0 counts and $n_1$ is the number of pixels with 1 count
(e.g., A03).
By ignoring all pixels with more than 1~count, this technique
is robust against background contamination from sources. 
The principal
requirement for using this Poisson-model technique is that the 
background counts are
low and follow a Poisson distribution; 
we show in \S4 that the background of the $\approx2$~Ms exposure 
meets this criterion.
We note that the background estimation is problematic for several sources
which are located close to bright sources or near the edge of the 
survey field
where there is a strong gradient in exposure time. For each of these sources,
we have measured its background counts in the background maps described in 
\S4, using an annulus outside of the source-extraction region. Note 
that when constructing the background maps, we filled in the masked 
regions with a local background assuming a probability distribution; 
thus small additional uncertainties could be introduced during this process and 
will be carried on to the background estimation here. 
There are 17 such sources and they are marked with ``B''
in column 49 of Table~2. The net number of source counts is calculated
by subtracting the background counts from the source counts.

For sources with fewer than 1000 full-band counts, we have chosen the aperture
radii based on the encircled-energy function of the \chandra\ PSF as determined
using the CXC's {\sc mkpsf} software \citep{Feigelson2000,Jerius2000}.
In the soft band, where the background is lowest, the aperture radius
was set to the 95\% encircled-energy radius of the PSF. In the full and 
hard bands, the
90\% encircled-energy radius of the PSF was used. Appropriate aperture
corrections were applied to the source counts by dividing the extracted source
counts by the encircled-energy fraction for which the counts were extracted.

For sources with more than 1000 full-band counts, systematic errors in the
aperture corrections often exceed the expected errors from photon statistics
when the apertures described in the previous paragraph are used. Therefore, for
such sources we used larger apertures to minimize the importance of the
aperture corrections; this is appropriate since these bright sources dominate
over the background. We set the aperture radii to be twice the 90\% 
encircled-energy full-band
radii and inspected these sources to verify that the
measurements were not contaminated by neighboring objects. No aperture
corrections were applied to these sources.

Manual correction of the source photometry was performed 
for sources having overlapping PSFs.
We manually separated 18 close doubles and 4 close triples, and these
sources are flagged with ``S'' in column 49 of Table~2.

We have performed several consistency tests to verify the quality of the
photometry. For example, we have checked that the sum of the counts measured in
the soft and hard bands does not differ from the counts measured in the full
band by an amount larger than that expected from measurement error. Systematic
errors that arise from differing full-band counts and soft-band plus hard-band
counts are estimated to be $\la4\%$.

When a source is not detected in a given band, an upper limit is calculated;
upper limits are indicated as a ``$-1.00$'' in the error columns. 
All upper limits are
determined using the circular apertures described above. When the number of
counts in the aperture is $\leq 10$, the upper limit is calculated using the
Bayesian method of \citet{Kraft1991} for 99\% confidence. The
uniform prior used by these authors results in fairly conservative upper limits
\citep[see][]{Bickel1992}, and other reasonable choices of priors 
do not materially
change our scientific results.  For larger numbers of counts in the aperture,
upper limits are calculated at the $3\sigma$ level for Gaussian statistics.

\item Columns~15 and 16: the right ascension and declination 
of the optical counterpart,
which was obtained by matching the \hbox{X-ray} source positions (columns~2 and
3) to WFI $R$-band source positions using a matching radius that is 1.5 times
the quadratic sum of 
the positional errors of the X-ray and optical sources (i.e.,
$r_{\rm m}=1.5\sqrt{\Delta_{\rm_X}^2+\Delta_{\rm o}^2}$).
This matching radius was chosen to provide a large number of 
optical counterparts without introducing too many false matches.
The WFI $R$-band observations have a 5$\sigma$ limiting AB magnitude of
$27.3$ over the entire \hbox{CDF-S} field.
For 4 sources (our sources ``74'', ``283'', ``328'', and ``431'') 
that have more
than one optical match, the magnitude difference 
between the counterparts is less than three in all cases, 
and therefore the source with the smallest offset was
selected as the most-probable counterpart.
Using these criteria, 344
($\approx74\%$) of the sources have optical counterparts. 
Sources with no
optical counterparts have these right ascension and declination values set to 
\hbox{``00 00 00.00''} and \hbox{``$-$00 00 00.0''}. 
We tested the reliability of the matching
by shifting the \hbox{X-ray} source coordinates
and recorrelating with the optical sources.
The matching is reliable (false-match probability $\la 8\%$) to        
$R\approx24$. The false-match probability rises to $\approx18\%$,
$\approx27\%$, and $\approx35\%$ at $R\approx25$, 26, and 27, respectively.

\item Column~17: the measured offset between the optical and \hbox{X-ray}
sources in arcseconds. Sources with no optical counterparts have
a value set to ``$-1.00$''. The offsets for all matches are below $2\farcs0$.

\item Column~18: the $R$-band AB magnitude of the optical counterpart.
Sources with no optical counterparts have a value set to ``$-1.00$''.

\item Columns~19 and 20: the corresponding source number and 
$i$-band AB magnitude from the GOODS-S
v2.0 $i$-band source catalog.\footnote{See \citet{Giavalisco2004} and
http://archive.stsci.edu/pub/hlsp/goods/catalog\_r2/.}
We matched the positions of the optical counterparts (see columns~15 
and 16) to the GOODS-S source positions using
a matching radius of $0\farcs5$. 
In 6 cases (our sources ``88'', ``120'', 
``135'', ``155'', ``313'', and ``322'') 
where there is more than one GOODS-S source matching to an
optical counterpart,
we selected the GOODS-S source with the smallest offset 
as the most-probable match. 218 matches were found for the 344 
optical counterparts; note that the GOOD-S field does not cover the
whole \hbox{CDF-S}.
By shifting the coordinates of the optical counterparts
and recorrelating with the GOODS-S sources, we estimated the 
false-match probability to be
$\la 5\%$.
The GOODS-S $i$-band observations have a 5$\sigma$ limiting AB magnitude of
$28.5$.
The $i$-band magnitude is the SExtractor \citep{Bertin1996}
corrected isophotal magnitude.
Sources with no GOODS-S match have 
these two columns set to ``$-1$'' and ``$-1.00$'', respectively.

\item Columns~21 and 22: the corresponding coordinate-based source 
name and $z$-band AB magnitude from the
Galaxy Evolution from Morphologies and SEDs (GEMS)
source catalog \citep{Caldwell2008}.
We matched the positions of the optical counterparts (see columns~15
and 16) to the GEMS source positions using
a matching radius of $0\farcs5$.
In 1 case (our source ``74'') where there is more than one 
GEMS source matching to an
optical counterpart,
we selected the GEMS source with the smallest offset
as the most-probable match. 297 matches were found for the 344
optical counterparts.
By shifting the coordinates of the optical counterparts
and recorrelating with the GEMS sources, we estimated the
false-match probability to be
$\la 2\%$. 
The GEMS $z$-band observations have a 5$\sigma$ limiting AB magnitude of
$27.3$ over the entire \hbox{CDF-S} field.
The $z$-band magnitude is the SExtractor MAG\_BEST magnitude.
Sources with no GEMS match have 
these two columns set to ``$-1$'' and ``$-1.00$'', respectively.

\item Columns~23 and 24: the corresponding source number and
$K_s$-band AB magnitude from the source catalog for the ESO/NTT
SOFI survey of the \hbox{CDF-S} region.\footnote{See 
http://www.eso.org/sci/activities/projects/eis/surveys/summary\_DPS.html.}
We matched the positions of the optical counterparts (see columns~15
and 16) to the SOFI source positions using
a matching radius of $0\farcs75$.
266 matches were found for the 344
optical counterparts.
By shifting the coordinates of the optical counterparts
and recorrelating with the SOFI sources, we estimated the
false-match probability to be
\hbox{$\la1\%$}.
The SOFI $K_s$-band observations have a 5$\sigma$ limiting AB magnitude of
$23.0$ over the entire \hbox{CDF-S} field. 
The $K_s$-band magnitude is the SExtractor corrected isophotal magnitude.
Sources with no SOFI match have  
these two columns set to ``$-1$'' and ``$-1.00$'', respectively.

\item Columns~25 and 26: the corresponding source number and IRAC
$5.8$ $\mu$m flux density ($f_{58}$) from the {\it Spitzer} 
IRAC/MUSYC Public Legacy Survey in the \hbox{E-CDF-S} (SIMPLE)
source catalog.\footnote{See http://ssc.spitzer.caltech.edu/legacy/simplehistory.html.}
We matched the positions of the optical counterparts (see columns~15
and 16) to the SIMPLE source positions using
a matching radius of $0\farcs75$.
306 matches were found for the 344
optical counterparts.
By shifting the coordinates of the optical counterparts
and recorrelating with the SIMPLE sources, we estimated the
false-match probability to be
$\la 2\%$. 
The SIMPLE $5.8$ $\mu$m observations have a 5$\sigma$ limiting AB magnitude of
21.9--22.5 over the entire \hbox{CDF-S} field; the limiting magnitude
is spatially dependent for SIMPLE.
The $5.8$ $\mu$m flux density is the aperture flux density
in a $2\farcs0$ circular aperture,
normalized to an AB magnitude zero point of 25. Note that 
an aperture correction of $\approx1.5$ was not applied to these fluxes;
i.e., the aperture-corrected AB magnitude is 
$m({\rm AB})=25-2.5\log10(1.5\times f_{58})$.
Sources with no SIMPLE match have
these two columns set to ``$-1$'' and ``$-1.00$'', respectively.

\item Columns~27 and 28: the corresponding spectroscopic redshift
and the reference for the redshift.
Secure spectroscopic redshifts were collected from
\citet{LeFevre2004}, \citet{Szokoly2004}, \citet{Mignoli2005},
\citet{Ravikumar2007}, \citet{Popesso2008}, and \citet{Vanzella2008}, 
with the reference numbers of 1--6 in column~28,
respectively. A matching radius of $0\farcs5$ was used when matching the
optical counterparts (see columns~15
and 16) to the redshift catalogs.
190 of the 344 optical counterparts have redshift measurements.
By shifting the coordinates of the optical counterparts
and recorrelating with the redshift catalogs, we estimated the
false-match probability to be
$\la 1\%$.
Sources with no secure spectroscopic redshift have
these two columns set to ``$-1.000$'' and ``$-1$'', respectively.
Note that there are also photometric redshifts available in the literature
\citep[e.g.,][]{Mobasher2004,Wolf2004}, but these are
not included in our catalogs.

\item Column~29: the corresponding $\approx$1~Ms \hbox{CDF-S} source 
number from the main
\chandra\ catalog presented in A03 (see column~1 of Table~A2a in A03). We 
matched our \hbox{X-ray} source positions to A03 source positions using 
a matching radius that is the
quadratic sum of
the $\approx3\sigma$ positional errors of the \hbox{CDF-S} 
and A03 \hbox{X-ray} sources. 
The $3\sigma$ positional error of a \hbox{CDF-S} 
source is approximately
twice the positional error quoted in column~4 (i.e., $2\Delta_{\rm_X}$), 
and that of an A03  
source is approximately
twice the positional error quoted in Table~A2a of A03.
The false match probability is less than 1\% with this matching radius.
Only one A03 match was found for each matched source.
In one case where two close-double sources matched to one A03 source, we 
chose the source with the smallest offset (source ``433'') 
as the most-probable match. 
We manually set the counterpart of the source with 
source number ``437''
to be source ``312'' in A03, 
because A03 apparently underestimated the positional error of this source.
Sources with no A03 match have a value of ``$-1$''.

\item Columns~30 and 31: the right ascension and declination of 
the corresponding 
A03 source indicated in column~29. Sources with no A03 match have right 
ascension and
declination values set to \hbox{``00 00 00.00''} and \hbox{``$-$00 00 00.0''}.

\item Columns~32 and 33: the corresponding $\approx$1~Ms \hbox{CDF-S} 
source ``ID'' number and ``XID'' number from the main
\chandra\ catalog presented in G02.
When matching our
\hbox{CDF-S} source positions with G02 counterparts, we removed offsets
to the G02 positions of $-1\farcs2$ in right ascension and $+0\farcs8$ in 
declination (see $\S$A3
of A03); these positions are corrected in the quoted source positions in
columns~34 and 35. We used a matching radius that is the
quadratic sum of 
the $\approx3\sigma$ positional errors of the \hbox{CDF-S} 
and G02 \hbox{X-ray} sources. 
The $3\sigma$ positional error of a \hbox{CDF-S}
source is approximately
twice the positional error quoted in column~4, and that of a G02
source is quoted in Table~2 of G02.
Only one G02 match was found for 
each matched source.
In three cases where two close-double sources matched to one G02 source, we
chose the source with the smallest offset 
(sources ``142'', ``195'' and ``275'') as the most-probable match.
Sources with no G02 match have a value of ``$-1$''.

\item Columns~34 and 35: the right ascension and declination of 
the corresponding G02 
source indicated in columns~32 and 33. Note that the quoted positions have
been corrected by the offsets described in columns~32 and 33 (see $\S$A3 of
A03). Sources with no G02 match have right ascension and declination 
values set to \hbox{``00 00
00.00''} and \hbox{``$-$00 00 00.0''}.

\item Columns~36--38: the effective exposure times determined from the
standard-band exposure maps (see \S3.1 for details on the exposure maps).
Dividing the counts listed in columns~6--14 by the corresponding effective
exposures will provide vignetting-corrected and quantum-efficiency
degradation corrected count rates.

\item Columns~39--41: the band ratio, defined as the ratio of counts
between the hard and soft bands, and the corresponding upper and lower errors,
respectively. Quoted band ratios have been corrected for differential
vignetting between the hard band and soft band using the appropriate exposure
maps. Errors for this quantity are calculated following the ``numerical
method'' described in \S1.7.3 of \citet{Lyons1991}; this avoids 
the failure of the
standard approximate variance formula when the number of counts is small (see
\S2.4.5 of \citealt{Eadie1971}). Note that the error distribution is not
Gaussian when the number of counts is small. Upper limits are calculated for
sources detected in the soft band but not the hard band, and lower limits are
calculated for sources detected in the hard band but not the soft band. For
these sources, the upper and lower errors are set to the computed band ratio.
Sources detected only
in the full band have band ratios and corresponding errors set to ``$-1.00$''.

\item Columns~42--44: the effective photon index ($\Gamma$) with upper and
lower errors, respectively, for a power-law model with the Galactic column
density given in \S1. When the number of source counts is not low, 
the effective photon index has been calculated based on the band ratio
in column~39 using the CXC's Portable, Interactive, Multi-Mission Simulator 
(PIMMS). Upper limits are calculated for
sources detected in the hard band but not the soft band, and lower limits are
calculated for sources detected in the soft band but not the hard band. For
sources with only limits on the effective photon index, 
the upper and lower errors are set to the computed effective
photon index.

A source with a low number of counts is defined as being (1) detected in the
soft band with $<30$ counts and not detected in the hard band, (2) detected in
the hard band with $<15$ counts and not detected in the soft band, (3) detected
in both the soft and hard bands, but with $<15$ counts in each, or (4) detected
only in the full band.  When the number of counts is low, the photon index is
poorly constrained and is set to $\Gamma=1.4$, a representative value for faint
sources that should yield reasonable fluxes. In this case, the upper and lower 
errors are set to ``$0.00$''.

\item Columns~45--47: observed-frame fluxes in the three standard bands;
quoted fluxes are in units of \flux.
Fluxes have been computed using the counts in columns~6--14, the
appropriate exposure maps (columns~36--38), and the effective photon indices 
given in column~42. The
fluxes have not been corrected for absorption by the Galaxy or material
intrinsic to the source. For a power-law model with $\Gamma=1.4$, the soft-band
and hard-band Galactic absorption corrections are $\approx$2.1\% and $\approx
0.1$\%, respectively. More accurate fluxes for these sources would require
direct fitting of the \hbox{X-ray} spectra for each observation, which is
model dependent and beyond the scope of this paper.

\item Column~48: the logarithm of the minimum false-positive probability
run with {\sc wavdetect} in which each source was detected (see $\S$3.2).  A
lower false-positive probability indicates a more significant source detection.
398 ($\approx86\%$) and 357 ($\approx77\%$) of our
sources are detected with false-positive probability thresholds of
1~$\times$~10$^{-7}$ and 1~$\times$~10$^{-8}$, respectively.

\item Column~49: notes on the sources. ``E'' refers to sources at the edge
that lie partially outside of the survey area.
 ``S'' refers to 
close doubles or triples where manual separation was required. 
``B'' refers to sources with background counts estimated using the 
background maps (see columns~6--14 of Table~2).

\end{enumerate}

In Table~3 we summarize
the source detections in the three standard bands.  
In total 462 point sources
are detected, 327 of which
were present in the main {\it Chandra} catalogs for the
$\approx$1~Ms \hbox{CDF-S} (G02 and A03), and thus 
135 sources are new.
For the 308 sources that were detected in the main catalog of A03,
we find general agreement between the derived \hbox{X-ray}
properties presented here and in A03. For example, we have compared the
full-band count rates of these 308 sources between the two catalogs.
The median ratio of the count rates is $\approx0.98$
with an interquartile range of $\approx$0.85--0.12.
Furthermore, the approximately doubled exposure improves the 
source positions and spectral constraints significantly, and thus the 
$\approx$2~Ms \hbox{CDF-S} catalogs presented here supersede those in A03.

Eighteen of the 326 sources detected in the 
main catalog of A03
are undetected here. Nine of these were detected in
{\sc wavdetect} runs with a false-positive probability threshold of
$1\times 10^{-5}$ in the present analysis. 
The other nine sources were weakly detected in A03
with less than 17 full-band counts. We examined the
regions of these nine sources in the three $\approx2$~Ms images and found no
emission clearly distinct from the background. 
Ten of the eighteen sources have optical counterparts in the
WFI $R$-band source catalog within $1\farcs3$, and three of them
are present in the
supplementary optically bright \chandra\ catalog (see \S3.3.3), suggesting
that they are likely true \hbox{X-ray} sources.
As the second $\approx1$~Ms
exposure was taken $\approx7$ years later, these eighteen sources could be
below our detection limit due to source variability or background 
fluctuations. A 30\% median flux variability has been observed for sources 
in the first $\approx1$~Ms data set \citep{Paolillo2004}, which is expected
to increase here owning to the long observation interval. 
There is also the possibility that some of the missing sources were false
detections in A03, since $\approx$3--9 false detections were expected (A03).

Four of the 304 sources in the main catalog of G02 are not detected here, 
two of which were detected in
{\sc wavdetect} runs with a false-positive probability threshold of
$1\times 10^{-5}$. All four sources lie at large off-axis angles, and none of 
them is in the A03 main catalog. These sources could be
below our detection limit due to source variability or background
fluctuations. Note that 19 G02 sources that were not detected in A03 
are detected here, suggesting that these are likely true sources. 
These sources were probably not reported in the A03 main catalog
due to the conservative {\sc wavdetect} false-positive 
probability threshold ($1\times 10^{-7}$) adopted in that work.

\begin{figure}
\centerline{\includegraphics[scale=0.5]{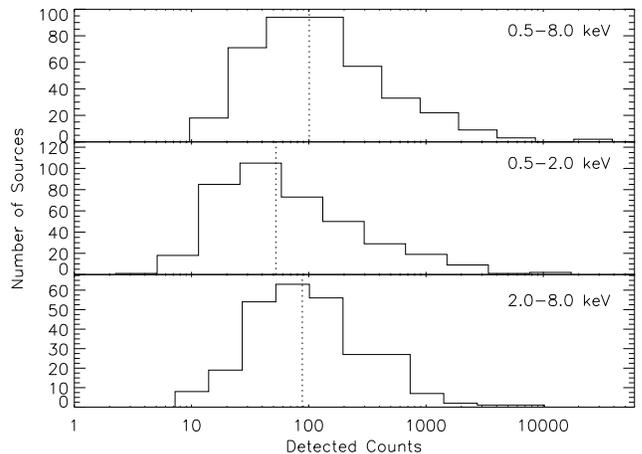}
} \figcaption{Histograms showing the distributions of detected source 
counts for sources in the main
{\it Chandra} catalog in the full ({\it top}), soft ({\it middle}), and 
hard ({\it
bottom}) bands. Sources with upper limits have not been included in these
diagrams.
The vertical dotted lines indicate median numbers of counts in each band
(see Table~3).
\label{cnthist}}
\end{figure}

\begin{figure}
\centerline{\includegraphics[scale=0.5]{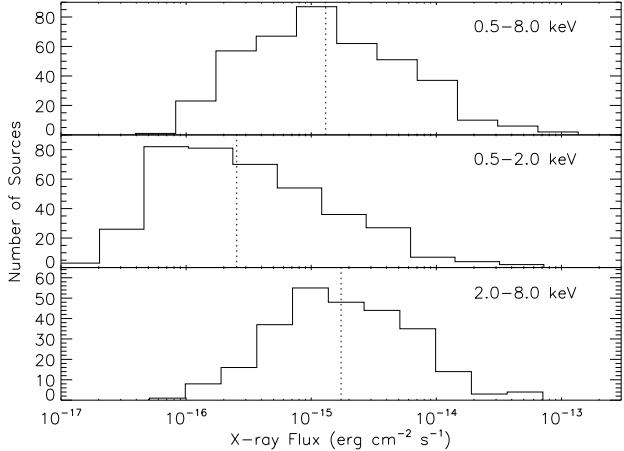}} 
\figcaption{Histograms showing the distributions of \hbox{X-ray} fluxes for 
sources in the main
{\it Chandra} catalog in the full ({\it top}), soft ({\it middle}), 
and hard ({\it
bottom}) bands. Sources with upper limits have not been included in this figure.
The vertical dotted lines indicate the median fluxes of 
$1.3\times$~10$^{-15}$, $2.5\times$~10$^{-16}$ and $1.7\times$~10$^{-15}$~\flux\
for the full, soft, and hard bands, respectively.
\label{fluxhist}}
\end{figure}

In Table~4 we summarize 
the number of sources detected in one band but not
another. There are three sources detected only in the hard band. 
For comparison, there is one source in the $\approx1$~Ms \hbox{CDF-S}
that was detected only in the hard band (A03).
In Figure~\ref{cnthist} we show the distributions of detected 
counts in the three standard
bands. The median numbers of counts for the full band, soft band and hard 
band are $\approx101$, $\approx53$ and $\approx89$, respectively.
There are 202 sources with $>100$ full-band counts, for which basic
spectral analyses are possible, and 33 sources with $>1000$ full-band
counts. 
In Figure~\ref{fluxhist} we show the distributions of \hbox{X-ray} flux 
in the three standard
bands. The \hbox{X-ray} fluxes in this survey span roughly four orders of
magnitude, with $\approx$50\% of the sources having soft-band and hard-band
fluxes of less than $2.5\times 10^{-16}$ \flux\ and 
$1.7\times 10^{-15}$ \flux, respectively.

\begin{figure}
\centerline{\includegraphics[scale=0.45]{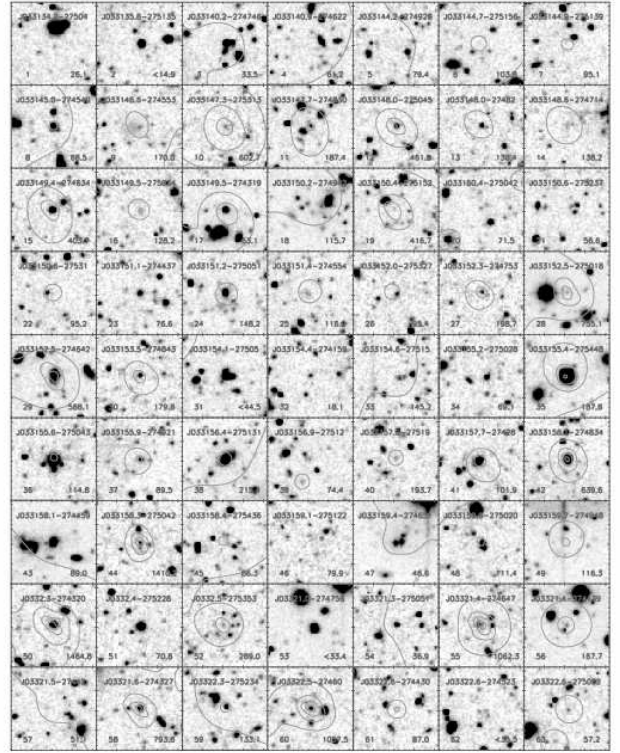}}
\figcaption{WFI $R$-band postage-stamp images for the sources in the main {\it Chandra}
catalog with full-band adaptively smoothed \hbox{X-ray} contours overlaid. The
contours are logarithmic in scale and range from \hbox{$\approx$0.003\%--30\%} 
of the maximum pixel value. The label at the top of each image gives the 
source name, which is composed of
the source coordinates, while numbers at the bottom left and right-hand corners
correspond to the source number (see column~1 of Table~2) and the full-band 
counts or upper limits (with a ``$<$'' sign) on the full-band counts, 
respectively. 
In several cases no \hbox{X-ray} contours are 
present, either because these sources were not detected in the full band or 
the full-band counts are low and {\sc
csmooth} has suppressed the observable emission in the adaptively smoothed
images. Each image is $25\arcsec$ on a side, and the
source of interest is always located at the center of the image. Only one of
the 8 pages of cutouts is included here; all 8 pages are
available in the electronic edition.
\label{ps}}
\end{figure}

\begin{figure*}
\centerline{\includegraphics[scale=0.4]{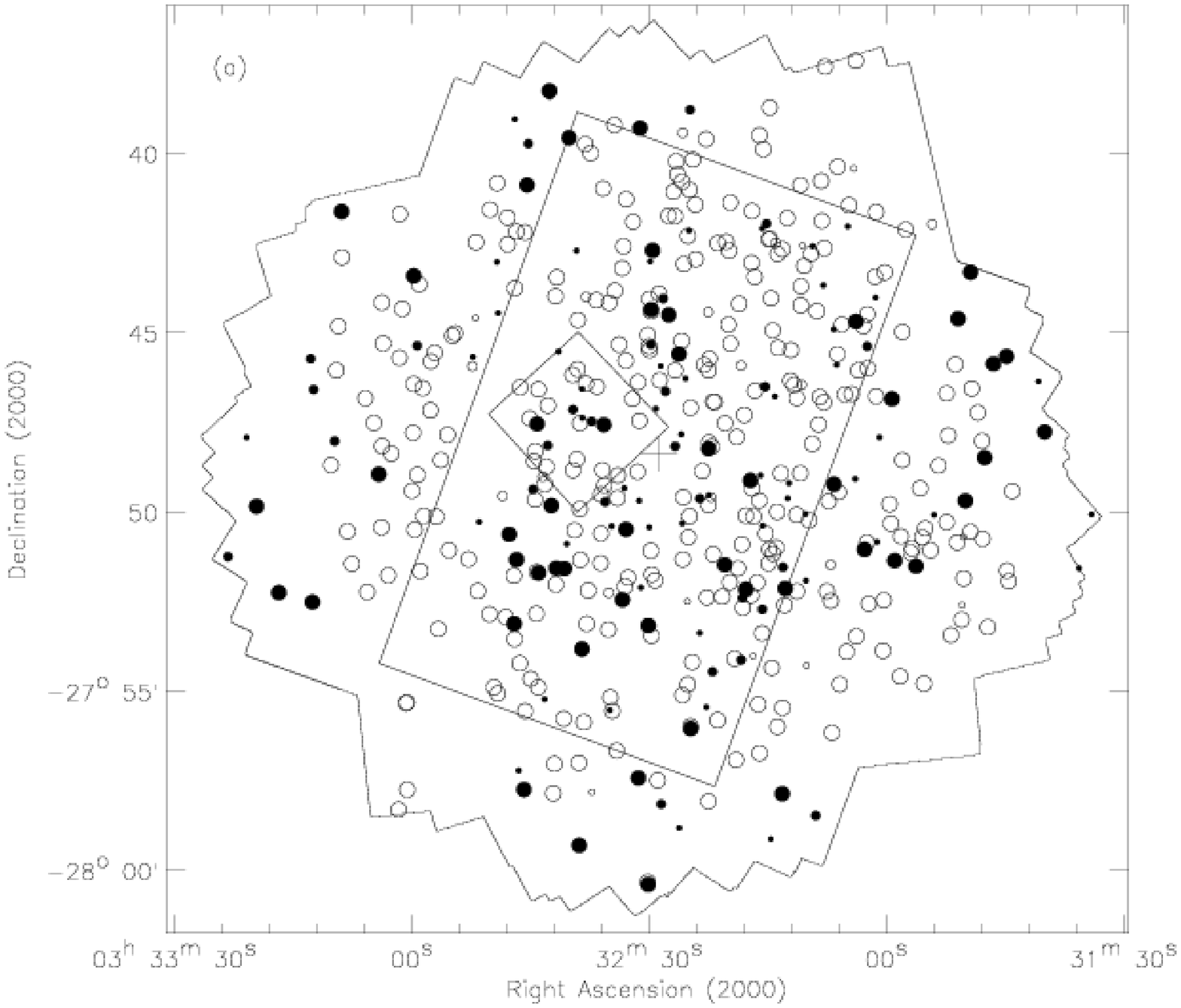}
\includegraphics[scale=0.4]{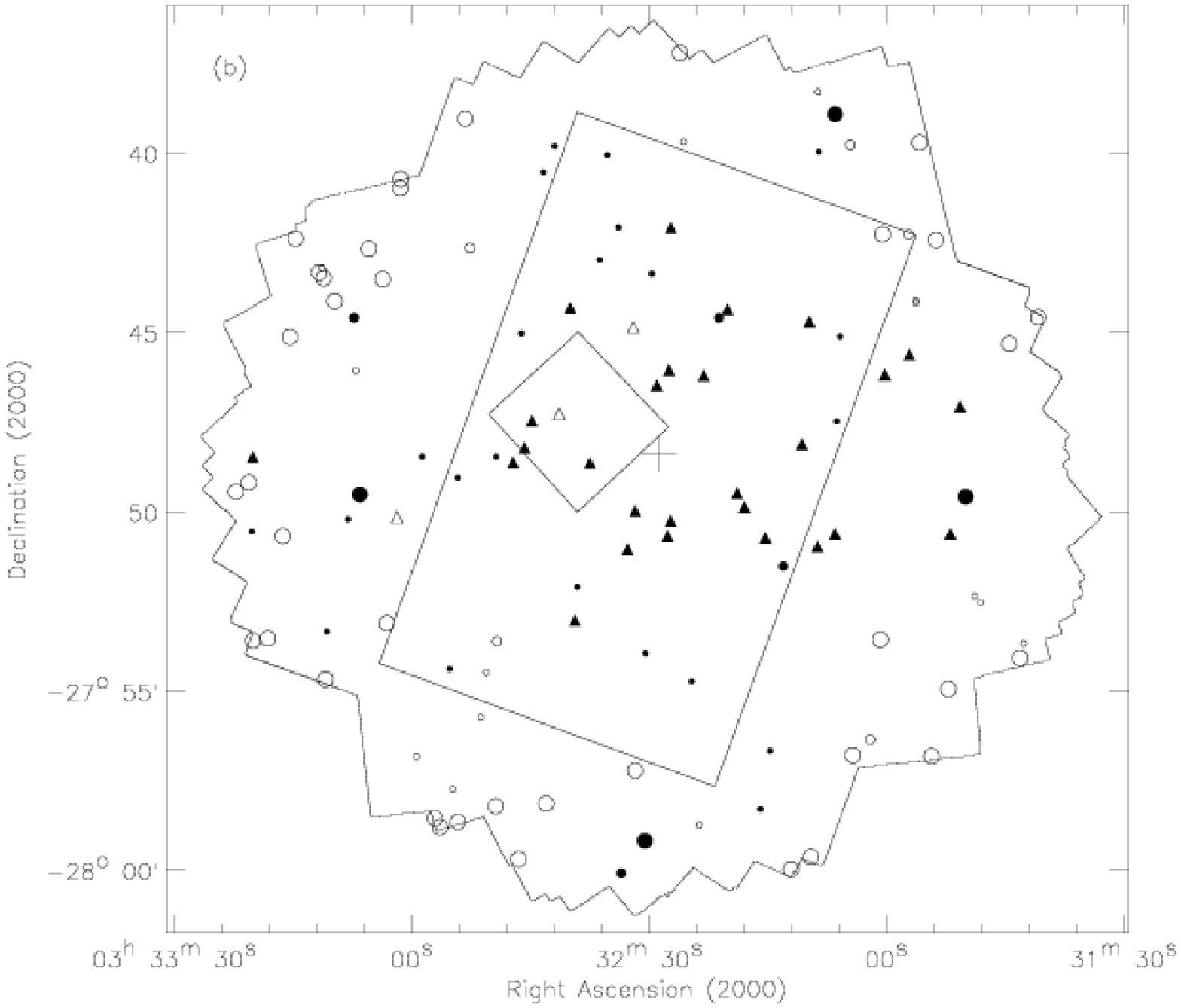}}
\figcaption{
Positions of the sources in ({\it a}) the main {\it Chandra} catalog
and ({\it b}) the supplementary {\it Chandra} catalogs.
Circles represent \hbox{X-ray} sources in ({\it a}) the main \chandra\
catalog and ({\it b}) the supplementary CDF-S plus E-CDF-S \chandra\ catalog.
Open circles represent \hbox{X-ray} sources that were
previously detected in ({\it a}) the main catalogs of G02 or A03
and ({\it b}) the main catalogs of G02, A03, or L05. Filled circles
represent new sources.
Sizes indicate the maximum detection significance corresponding to
{\sc wavdetect} false-positive probability detection thresholds of 
$1\times10^{-8}$ ({\it large circles}),
$1\times10^{-7}$ ({\it medium circles}), and $1\times10^{-6}$ 
({\it small circles}).
Sources in the optically bright catalog
are shown as open triangles (previously detected in the main catalog of
A03) and filled triangles (new sources) in ({\it b}).
For sources in the CDF-S plus E-CDF-S catalog, 
their detection significances are preferentially higher near the edge of 
the field due to the contribution of the
\hbox{E-CDF-S} exposure.
The regions and the cross symbol have the same meaning as those in
Fig. \ref{fbimg}.
\label{pos}}
\end{figure*}

In Figure~\ref{ps} we show ``postage-stamp'' images from the WFI 
$R$-band image with
adaptively smoothed full-band contours overlaid for sources in the
main {\it Chandra} catalog. The wide range of \hbox{X-ray} source sizes
observed in these images is largely due to PSF broadening with off-axis angle.
Figure~\ref{pos}{\it a} shows the positions of sources detected in the main {\it
Chandra} catalog. The source density is highest 
close to the average aim point where the sensitivity is highest. 
Different symbol sizes represent different significances of
source detection with {\sc wavdetect} (see column~48 of Table~2).
New \hbox{X-ray} sources that are not present in the 
G02 or A03 main catalogs are indicated as filled circles; 135 new 
sources are detected, of which 15 lie outside the solid-angle 
coverage of the first $\approx1$~Ms exposure.


\begin{figure}
\centerline{\includegraphics[scale=0.5]{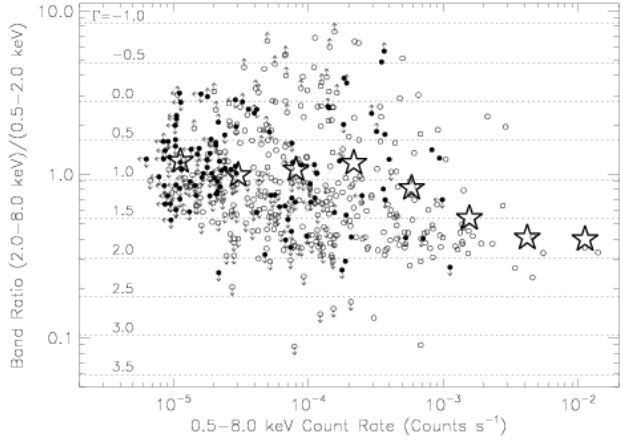}
}
\figcaption{
Band ratio vs. full-band count rate for sources in the main \chandra\ 
catalog. Open circles represent \hbox{X-ray} sources that were
detected in the main catalogs of G02 or A03.
Filled circles represent new sources.
Plain arrows indicate upper or lower limits. Sources detected only
in the full band cannot be plotted. 
The open stars show average band ratios as a function of full-band
count rate derived from stacking analyses.
Horizontal dotted lines show the band ratios 
corresponding to given effective photon indices; these were 
calculated using PIMMS.
\label{bratio}}
\end{figure}

\begin{figure*}
\centerline{\includegraphics[scale=0.5]{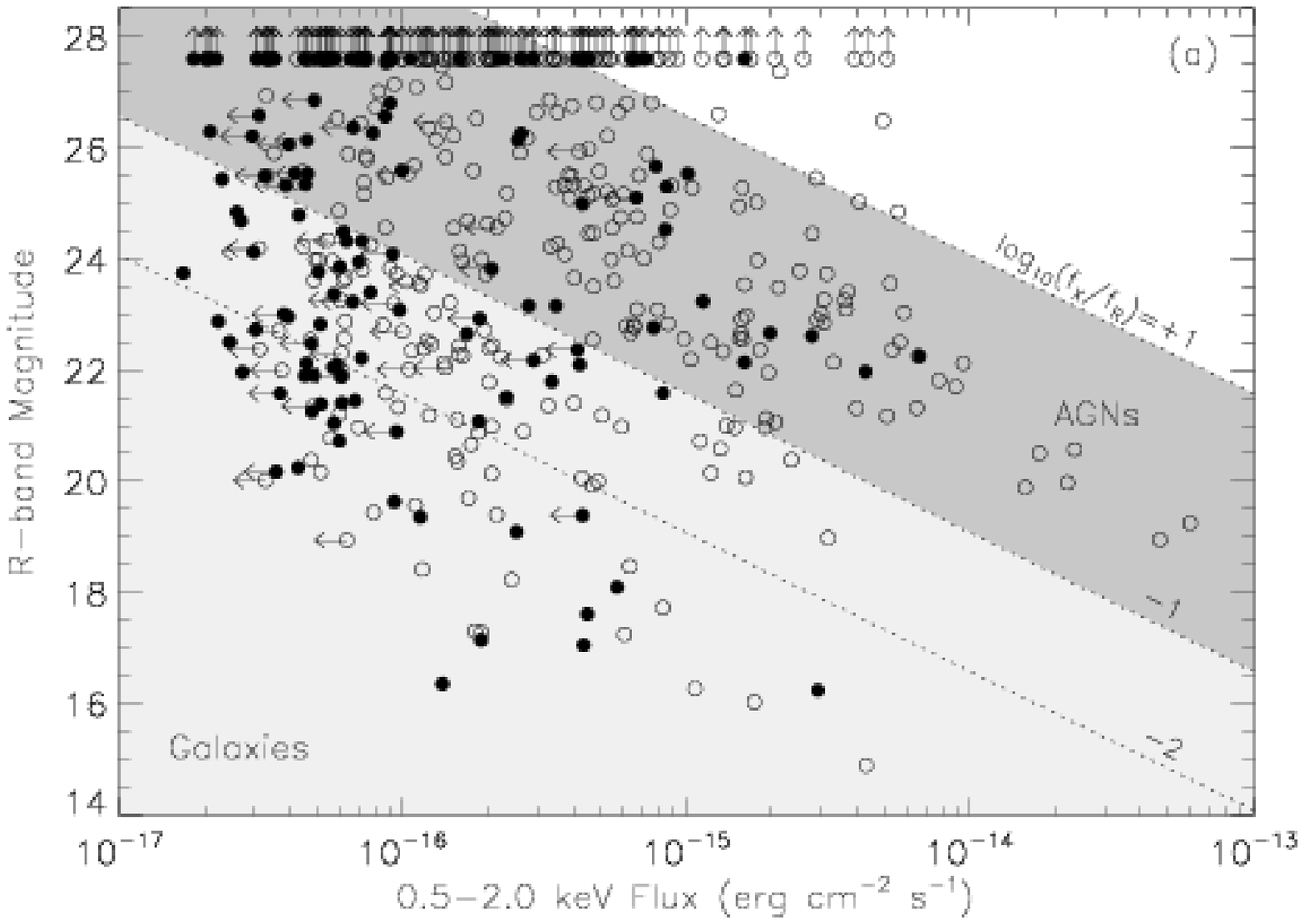}
\includegraphics[scale=0.5]{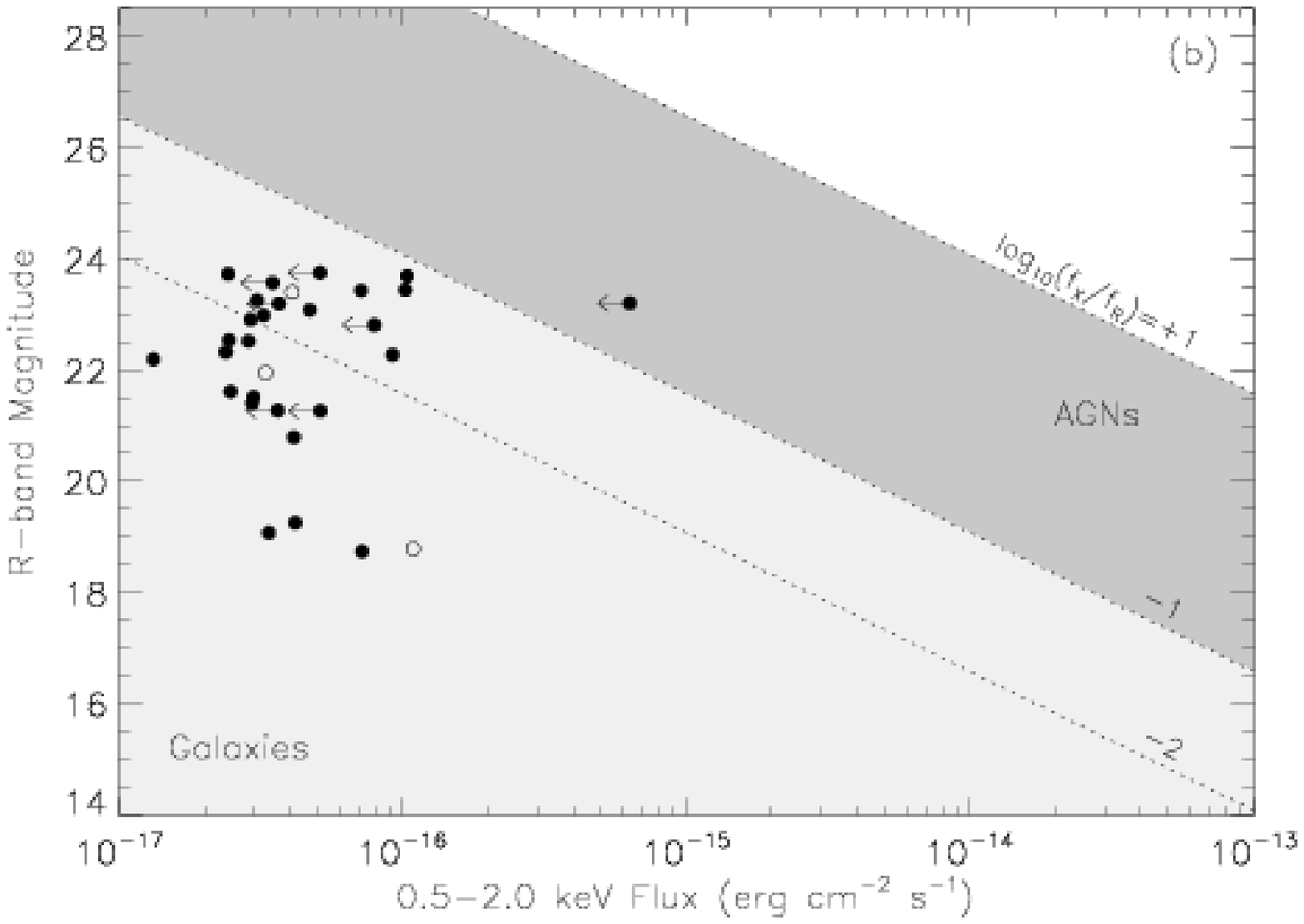}
}
\figcaption{
WFI $R$-band magnitude vs. soft-band flux for \hbox{X-ray} sources
in ({\it a}) the main catalog 
and ({\it b}) the supplementary optically bright catalog. 
Open circles represent \hbox{X-ray} sources that were
detected in the main catalogs of G02 or A03.
Filled circles represent new sources.
Sources without an optical counterpart are plotted as upward arrows.
Diagonal lines indicate constant flux ratios. The shaded areas show the 
approximate flux ratios for AGNs (dark gray) and galaxies (light gray).
\label{fox}}
\end{figure*}

Figure~\ref{bratio} shows the band ratio as a function of full-band count
rate for sources in the main \chandra\ catalog. 
We also derived average band ratios by stacking the individual 
sources together using a procedure similar to that of 
\citet{Lehmer2008}.
The average
band ratio rises at lower count rates. The corresponding 
average photon index flattens from $\Gamma\approx1.8$ to 
$\Gamma\approx0.8$ for full-band count rates of $\approx10^{-2}$
to $\approx2\times10^{-4}$ counts s$^{-1}$.
This trend has been
reported in other studies (e.g., \citealt{Tozzi2001}; A03; L05) and 
is due to an increase in the number of absorbed AGNs 
detected at fainter fluxes. 
The average photon index does not continue getting flatter 
below full-band count rates of
$\approx2\times10^{-4}$ counts s$^{-1}$, probably due to the increased 
contribution from normal and starburst galaxies at these lowest count rates
\citep{Bauer2004}.
In Figure~\ref{fox}{\it a} we show the WFI $R$-band magnitude versus 
soft-band flux for \hbox{X-ray} sources
in the main catalog, as well as  
the approximate flux ratios for AGNs and galaxies 
\citep[e.g.,][]{Maccacaro1988,Stocke1991,Hornschemeier2001,Bauer2004}.
More than half (304) of the \hbox{X-ray}
sources lie in the region expected for AGNs, 74 of which are new sources.
A significant minority (158) of the sources lie in the region for normal and 
starburst galaxies, 61 of which are new sources. The new sources 
have an increased fraction of normal and starburst galaxies.
This source characterization, based only on the X-ray--to--optical flux 
ratio, is only approximate and will be refined in future studies.


\subsubsection{Supplementary CDF-S plus E-CDF-S Chandra Source Catalog}

We can gain additional sensitivity in the outer
portions of the $\approx2$~Ms CDF-S footprint by including 
the $\approx250$~ks \hbox{E-CDF-S} 
(L05) observations. To this end, we processed and registered 
the \hbox{E-CDF-S} exposures in the same manner as 
our \hbox{CDF-S} observations. Notably, because of
the different coverage of the \hbox{CDF-S} and \hbox{E-CDF-S} 
(see Figure~2 of L05), the PSF sizes for the \hbox{E-CDF-S} near
the average aim point for the CDF-S are substantially
larger than those for the \hbox{CDF-S}.
The \hbox{E-CDF-S} will likely only contribute additional background for
all but the strongest sources around the center of the field. 
Thus, we excluded the \hbox{E-CDF-S} event
lists within 4$\arcmin$ of the \hbox{CDF-S} average aim point. 
We also masked out portions of the \hbox{E-CDF-S} where the \hbox{CDF-S}
exposure time was zero. Images and exposure maps were cropped in 
a similar manner.  

We ran {\sc wavdetect} with a false-positive probability
threshold of $1\times 10^{-6}$ on the three standard-band images 
for the combined \hbox{CDF-S} plus \hbox{E-CDF-S}, detecting 86
sources not present in the main \chandra\ source
catalog. 
The positions of these sources have been improved following the procedure
described in \S3.3.1.
Due to the drastically different overlapping PSFs, the derived 
properties of these
\hbox{X-ray} sources are not as reliable as those in the main catalog. 
Therefore
we present these sources in Table~\ref{tbl-sp1} as 
a supplementary CDF-S plus E-CDF-S 
\chandra\ source catalog. For sources already detected in 
the \hbox{E-CDF-S} (L05), we took the photometry data from L05 directly.
For new sources, photon counts and effective exposure times
were extracted separately from the \hbox{CDF-S} and \hbox{E-CDF-S} 
data sets and then summed to give a total
number of counts and a total effective exposure time.
The format of Table~\ref{tbl-sp1} is very
similar to that of Table~2, with a few details given below.

\begin{enumerate}

\item Columns~1--28: the format of these columns is exactly the same as 
that of columns~1--28 in Table~2, so the column descriptions in \S3.3.1 
are applicable.
Note that for sources detected in the \hbox{E-CDF-S} (see column~29 or 52), 
the source
counts and their uncertainties were taken from L05 directly.

\item Column~29: the corresponding $\approx$250~ks \hbox{E-CDF-S} source
number from the main
\chandra\ catalog presented in L05 (see column~1 of Table~2 in L05). We
matched our \hbox{X-ray} source positions to L05 source positions using
a matching radius that is the
quadratic sum of
the $\approx3\sigma$ positional errors of the \hbox{CDF-S}
and L05 \hbox{X-ray} sources.
The $3\sigma$ positional error of a \hbox{CDF-S}
source is approximately
twice the positional error quoted in column~4, and that of an L05
source is approximately
twice the positional error quoted in Table~2 of L05.
Only one L05 match was found for each matched source.
Sources with no L05 match have a value of ``$-1$''.

\item Columns~30 and 31: the right ascension and declination of
the corresponding
L05 source indicated in column~29. Sources with no L05 match have right
ascension and
declination values set to \hbox{``00 00 00.00''} and \hbox{``$-$00 00 00.0''}.

\item Columns~32--51: the format of these columns is exactly the same as
that of \hbox{columns~29--48} in Table~2, so the column descriptions in \S3.3.1 
are applicable. Note that for sources detected in the \hbox{E-CDF-S} 
(see column~29 or 52), the source exposure times, band ratios, photon indices,
and fluxes were taken from L05 directly.

\item Column~52: notes on the sources. ``L'' refers to sources that 
were detected in the $\approx$250~ks \hbox{E-CDF-S} (L05). 

\end{enumerate}

The 86 CDF-S plus E-CDF-S sources have 
effective exposures up to $\approx1.9$~Ms. 
Their
positional uncertainties were estimated following equation (1),
though the positional accuracy of the off-axis sources will often have been
improved due to the small PSF sizes of the \hbox{E-CDF-S}. 60 ($\approx70\%$)
of the sources have optical counterparts. Two of the 86 sources have 
counterparts in 
the A03 main catalog and another two have counterparts 
in the G02 main catalog. In addition, 53 of the sources were detected in 
the main catalog of L05. There are thus 30 new sources in this 
supplementary catalog.
50 ($\approx57\%$) and 41 ($\approx47\%$) of these
sources are detected with false-positive probability thresholds of
1~$\times$~10$^{-7}$ and 1~$\times$~10$^{-8}$, respectively.

Figure~\ref{pos}{\it b} shows the positions of sources detected in the 
supplementary \hbox{CDF-S} plus \hbox{E-CDF-S} catalog. 
Different symbol sizes represent different significances of the
source detection with {\sc wavdetect} (see column~51 of Table~\ref{tbl-sp1}).

\subsubsection{Supplementary Optically Bright Chandra Source Catalog}

Since the density of optically bright sources on the sky is comparatively
low, we constructed a supplementary \chandra\ source catalog including 
\hbox{X-ray} sources detected at a lower \hbox{X-ray} significance 
threshold than that used in the main catalog and having bright optical
counterparts. We ran {\sc wavdetect} with a false-positive probability
threshold of $1\times 10^{-5}$ on the three \hbox{CDF-S} images, 
and we found 132 lower
significance \hbox{X-ray} sources not present in the main \chandra\ source
catalog or the supplementary \hbox{CDF-S} plus \hbox{E-CDF-S}
catalog.
 
Bright optical sources were selected from the WFI $R$-band source 
catalog described in \S3.1, with an $R$-band magnitude brighter than 
23.8. This $R$-band cutoff was empirically determined to provide a good 
balance between the number of detected sources and the
expected number of false sources. We
searched for bright optical counterparts to the low-significance 
\hbox{X-ray} sources using a matching radius of $1\farcs 3$.
A matching radius of $1\farcs3$ was chosen as a compromise between having too
few matches and too many false matches.
In total 30 optically bright \hbox{X-ray} sources were found. 
We estimated the expected number of false matches by
manually shifting the \hbox{X-ray} source coordinates in right
ascension and declination by $5\farcs0$ and $10\arcsec$ 
(both positive and negative
shifts) and recorrelating with the optical sources. On average, the
number of false matches is $\approx3$ ($\approx10\%$), demonstrating 
that the majority
of the 30 \hbox{X-ray} matches are real \hbox{X-ray} sources. 

The supplementary optically bright \chandra\ source catalog is presented in
Table~\ref{tbl-sp2}. These sources typically have 4--35 counts in the 
band in which they were detected.
The format of Table~\ref{tbl-sp2} is 
similar to that of Table~2, 
with the details of the columns given below.

\begin{enumerate}

\item
Column~1: the source number. Sources are listed in order of increasing
right ascension.

\item Columns~2 and 3: the right ascension and declination of 
the \hbox{X-ray} source,
respectively. The {\sc wavdetect} positions are used here for these faint
\hbox{X-ray} sources. Whenever possible, we have quoted
the position determined in
the full band; when a source is not detected in the full band, we used, in
order of priority, the soft-band position or hard-band position.

\item Column~4: the positional uncertainty. For these faint
\hbox{X-ray} sources, the positional uncertainty is set to $1\farcs2$, 
the approximate 90th percentile of the optical--X-ray positional offsets 
given in column~17.

\item Column~5: the off-axis angle of the \hbox{X-ray} source in arcminutes 
(see column~5 of Table~2 for details).

\item Columns~6--14: the source counts and the corresponding $1\sigma$
statistical errors \citep{Gehrels1986} or the upper limits on source counts
for the three standard bands, respectively.
When a source is detected in a given band, the photometry is taken 
directly from {\sc wavdetect}. When a source is not detected, an upper limit
is calculated (see columns~6--14 of Table~2 for details).

\item Columns~15 and 16: the right ascension and declination of the 
optical counterpart.

\item Column~17: the measured offset between the optical and \hbox{X-ray}
sources in arcseconds.

\item
Column~18: the $R$-band AB magnitude of the optical counterpart.

\item
Columns~19--26: the $i$, $z$, and $K_s$ band AB magnitudes and the IRAC
5.8 $\mu$m flux density of the optical counterpart, and the correspoding
source ID in the optical and infrared catalogs (see columns~19--26
of Table~2 for details).

\item
Columns~27 and 28: the corresponding spectroscopic redshift
and the reference for the redshift (see columns~27 and 28 of Table~2 
for details).

\item Column~29: the corresponding $\approx$1~Ms CDF-S source number 
from the main
\chandra\ catalog presented in A03 (see column~1 of Table~3a in A03).  
We used a
matching radius that is the quadratic sum of the $\approx3\sigma$ 
positional errors of the
\hbox{CDF-S} and A03 \hbox{X-ray} sources.  
The $3\sigma$ positional error of a \hbox{CDF-S}
source is $\approx1\farcs3$, and that of an A03
source is approximately
twice the positional error quoted in Table~A2a of A03.
Only one A03 match was found for each matched source. Supplementary sources 
with no A03 match have a value of ``$-1$''. There are no matches to the 
main source catalog in G02, so we do not list the match results in this
table.

\item Columns~30 and 31: the right ascension and declination of the 
corresponding A03 source indicated in column~29. Sources with no A03 match have right ascension and declination values set to \hbox{``00 00 00.00''} 
and \hbox{``$-$00 00 00.0''}.

\item Columns~32--34: the effective exposure times derived from the
standard-band exposure maps.

\item Column~35: the photon index used to calculate source fluxes
(columns~36--38). We used a constant photon index of $\Gamma=2.0$ since our
source-selection technique preferentially selects objects with flux-ratios
$f_{\rm 0.5-2.0~keV}/f_R < 0.1$, which are observed to have effective photon
indices of $\Gamma \approx 2$ (e.g., $\S$~4.1.1 of Bauer \etal~2004).

\item Column~36--38: observed-frame fluxes in the three standard bands;
quoted fluxes are in units of \flux\ and have been calculated assuming
$\Gamma=2.0$. The fluxes have not been corrected for absorption by the Galaxy
or material intrinsic to the sources (see \hbox{columns~45--47} of Table~2 for
details).

\end{enumerate}

The WFI $R$-band magnitudes of these supplementary sources 
span $R=$18.7--23.8.
In Figure~\ref{fox}{\it b} we show the $R$-band magnitude versus
soft-band flux for the 30 optically bright \hbox{X-ray} sources.
The approximate flux ratios for AGNs and galaxies are also plotted.
The majority of the sources have the X-ray--to--optical flux ratios expected
for normal and starburst galaxies.
Some of these sources may be low-luminosity AGNs; only one source is 
detected in the hard band, suggesting that they are
unlikely to be luminous absorbed AGNs.
Note that the supplementary optically bright sources are not 
representative of the faintest \hbox{X-ray} sources as a whole, because our
selection criteria preferentially select optically bright and \hbox{X-ray}
faint non-AGNs (e.g., A03; \citealt{Hornschemeier2003}).
The positions of the sources in the
supplementary optically bright catalog are shown in Figure~\ref{pos}{\it b}.

\section{BACKGROUND AND SENSITIVITY ANALYSIS}
Background maps were created for the three standard bands. We 
first masked out the point sources from the main \chandra\ catalog 
using apertures with radii twice
that of the $\approx$90\% PSF encircled-energy radii; approximately
12\% of the pixels were masked out. The resultant images
should include minimum contributions from detected point sources.
However, they will include contributions from a few extended sources 
\citep[e.g.,][]{Bauer2002}, 
which will cause a slight overestimation of the measured
background. 
Even with a $\approx2$~Ms exposure, about 79\%
of the pixels have no background counts in the full band.
For such a small number of detected counts per pixel, the 
expected counts distribution is Poissonian. 
We compared the background-count 
distributions to Poisson distributions with the mean number
of background counts per pixel using the Kolmogorov-Smirnov test, 
and we found them to be very similar in 
all three standard bands for various regions across the survey field
(see $\S$4.2 of A03 for more details on the tests).
We filled in the masked regions for each source with a local background
estimate by constructing a probability distribution of counts 
using an annulus with
inner and outer radii of 2 and 4 times the $\approx$90\% PSF encircled-energy
radius, respectively. 
The background properties are summarized in
Table~\ref{tbl-bkg}. The total background includes contributions
from the unresolved cosmic background, particle background, and instrumental
background \citep[e.g.,][]{Markevitch2001,Markevitch2003}. For our analyses we
are only interested in the total background and do not distinguish between
these different components.
The mean background count rates are $\approx20\%$--$30\%$ higher
compared to the $\approx2$~Ms \hbox{CDF-N} (A03) 
or the $\approx250$~ks \hbox{E-CDF-S} (L05), which are reasonable 
variations given the variability of the particle and instrumental
background components over the past several years. 

The faintest sources in the main \chandra\ catalog have $\approx5$
counts in the soft band and $\approx8$ counts in the hard band 
(see Table~\ref{tbldet}).
For a $\Gamma=$~1.4 power law with Galactic absorption, the corresponding
soft-band and hard-band fluxes at the average 
aim point are $\approx1.6\times10^{-17}$
\flux\ and $\approx9.0\times10^{-17}$ \flux,
respectively. This provides an estimate of the ultimate sensitivity of this 
survey.
However, these numbers are only relevant for a small area close to the 
average aim
point. To determine the sensitivity across the field it is necessary
to take into account the broadening of the PSF with off-axis angle, 
as well as
changes in the effective exposure and background rate across the field. 
Following L05, we
estimated the sensitivity across the field by employing a Poisson model,
The resulting relation can be approximately represented by

\begin{equation}
\log N~=\alpha + \beta \log b + \gamma (\log b)^2+ \delta (\log b)^3
\label{eqsen}
\end{equation}

\noindent where $N$ is the required number of counts for detection, and
$b$ is the number of background counts in a source cell; $\alpha=0.917$,
$\beta=0.414$, $\gamma=0.0822$, and $\delta=0.0051$ are fitting constants.
For the sensitivity calculations here, we measured the number of background 
counts $b$ in the background
maps using an aperture size of 70\% of the PSF encircled-energy radius. The 
70\% encircled-energy radius was chosen as a compromise between having too
few source counts and too many background counts. 

\begin{figure}
\centerline{\includegraphics[scale=0.4]{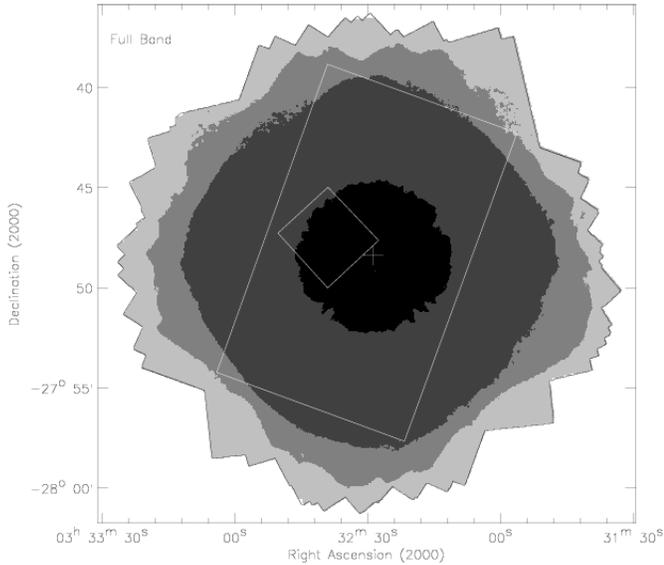}}
\figcaption{
Full-band sensitivity map of the 2~Ms \hbox{CDF-S}. This
sensitivity map has been created following $\S$4. 
The gray-scale levels (from
black to light gray) represent areas with flux limits (in units of \flux) 
of $<10^{-16}$, \hbox{$10^{-16}$--$3.3\times10^{-16}$}, \hbox{$3.3\times
10^{-16}$--$10^{-15}$}, and $>10^{-15}$, respectively. 
The regions and the cross symbol have the same meaning as those in
Fig.~\ref{fbimg}.
\label{senmap}}
\end{figure}

\begin{figure}
\centerline{\includegraphics[scale=0.5]{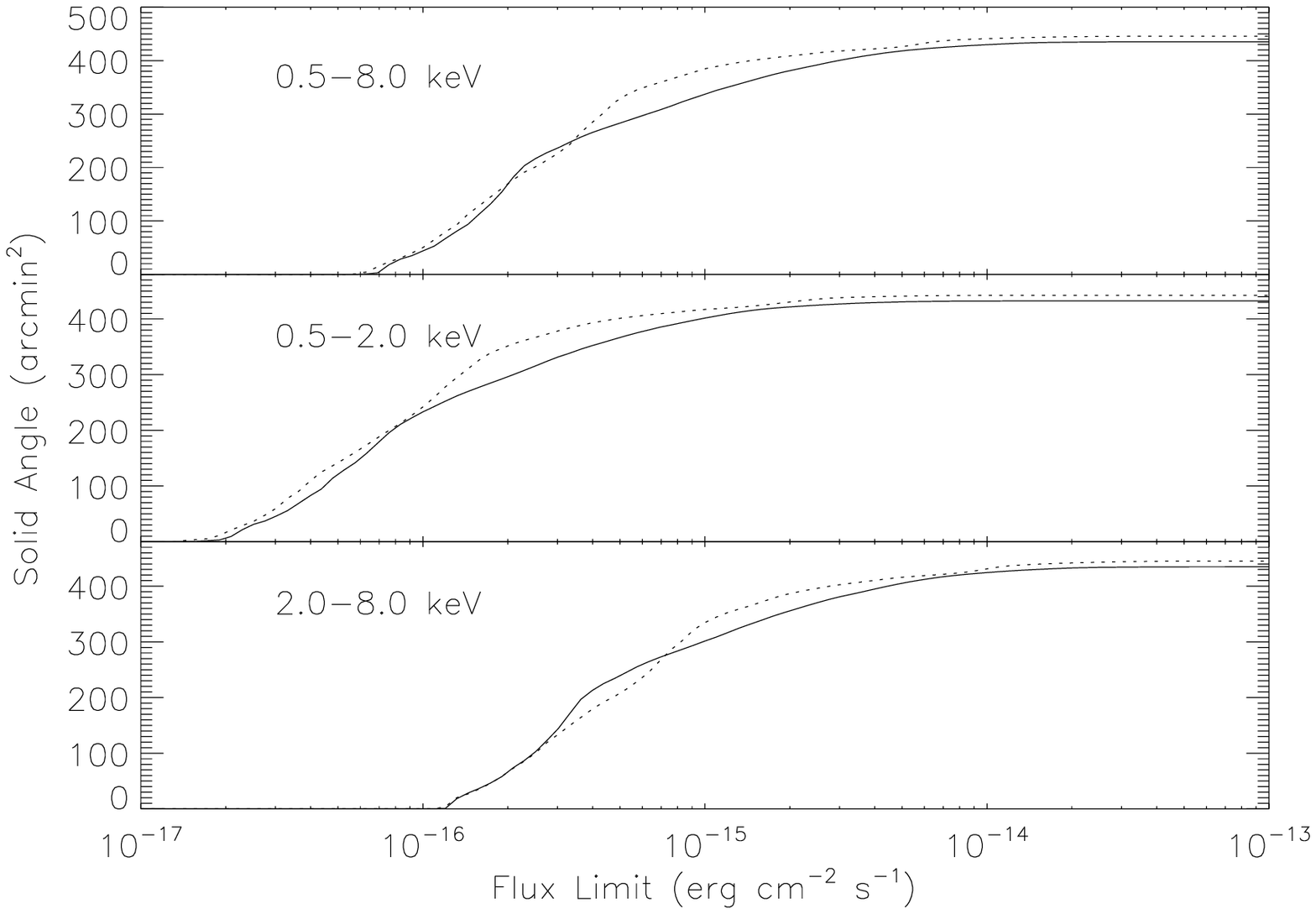}}
\figcaption{
Survey solid angle as a function of the flux limit for the 
full ({\it top}), soft
({\it middle}), and hard ({\it bottom}) bands, determined following $\S$4.
Data are plotted as solid curves for the $\approx2$~Ms \hbox{CDF-S}, 
and as dotted curves for the $\approx2$~Ms \hbox{CDF-N}.
The flux limits at the average aim point of the \hbox{CDF-S}
are $\approx7.1\times10^{-17}$
\flux\ (full band), $\approx1.9\times10^{-17}$ \flux\ (soft band), 
and $\approx1.3\times10^{-16}$ \flux\ (hard band).
\label{senhist}}
\end{figure}

Following equation (\ref{eqsen}), 
we constructed sensitivity maps using the background and
exposure maps, assuming a $\Gamma=$~1.4 power-law model with Galactic
absorption. Since we do not filter
out detected sources with our sensitivity maps, a small fraction 
of sources have
fluxes slightly below
these sensitivity limits (4 sources in the full band, 
14 sources in the soft band, and 7 sources in the
hard band). 
The full-band sensitivity map is shown in Figure~\ref{senmap},
and in
Figure~\ref{senhist} we show plots of solid angle versus flux limit 
for the full, soft,
and hard bands. The $\approx$1~arcmin$^2$ region at the average 
aim point has
soft-band and hard-band sensitivity limits of
$\approx1.9\times10^{-17}$ \flux\ and
$\approx1.3\times10^{-16}$ \flux, respectively. Solid angles for
the $\approx2$~Ms \hbox{CDF-N} have been plotted for comparison in
Figure~\ref{senhist} (dotted curves),
which appear to be similar to those for the \hbox{CDF-S}.\footnote{
The \hbox{CDF-N} sensitivity limits were calculated following the same method
described above.}


\section{NUMBER COUNTS FOR THE MAIN {\bf {\em CHANDRA}} CATALOG}
Cumulative number counts, $N(>S)$, for the soft and hard
bands were calculated for the $\approx2$~Ms \hbox{CDF-S}.
To quantify the effects of incompleteness and bias, 
we took a similar approach to the one in \citet{Bauer2004} and created
200 Monte Carlo simulated observations in both the soft and hard bands.
We added simulated sources at random positions to the background maps
described in \S4. The fluxes of these simulated sources were drawn randomly
from the total number-count models of \citet{Moretti2003} between
$1.6\times10^{-17}$ and $10^{-11}$ \flux\ in the soft band 
and $9\times10^{-17}$ and $10^{-11}$ \flux\ in the hard band. These
fluxes were converted to \hbox{X-ray} photon counts using the exposure maps
and a photon index of $\Gamma=1.4$. Statistical errors were added to 
the counts to account for the effect of Eddington bias. Finally, counts
for each simulated source were added to the background map following
a PSF probability distribution function derived from the combined model PSF
of the nearest real X-ray source in the main catalog. These model PSFs
were produced using AE.

Source searching and photometry were performed on the simulated images 
using the same method as that used to produce the main catalog. A 
completeness correction factor ($F$) was estimated by comparing the number 
of simulated input sources with the number of simulated detected sources as a 
function of detected counts. A flux recovery correction factor ($R$)
was calculated by comparing the simulated input counts with simulated measured
counts. The correction factors are position- and count-dependent. 
For each of the 462 X-ray sources in the main catalog, we determined
the two correction factors based on a sample of simulated 
sources within $2\arcmin$ of the source position and having similar exposure
times. 
Sources close to the edge of the survey field are not well sampled, and 
thus we calculated cumulative number
counts using only the 428 X-ray sources that are located
within $10\arcmin$ of the average aim point.
The completeness and flux 
recovery corrections remain close to unity above $\sim$50--100 counts.
Below this point, \chandra's varying PSF size and spatially dependent
vignetting begin to affect source detection and photometry.

We set our minimum flux levels to $3\times10^{-17}$ \flux\ in
the soft band and $2.5\times10^{-16}$ \flux\ in the hard band.
These limits were chosen since at lower fluxes there are less than 10--15
additional sources contributing to the number counts, and thus the 
number counts at fainter levels have large uncertainties.
The cumulative number of
sources, $N(>S)$, brighter than a given flux, $S$, weighted by the appropriate
aerial coverage, is
\begin{equation}
N(>S) = \sum_{S_i > S} (F_i\Omega_i)^{-1}~,
\end{equation}
\noindent where $\Omega_i$ is the maximum solid angle for which a source with
flux, $S_i$, could be detected. Each flux $S$ has been corrected for flux bias
assuming
\begin{equation}
S_i=R_iS_i^0~,
\end{equation}
\noindent where $S_i^0$ is the original flux quoted in the main catalog.
The maximum solid angles were computed
using the inner $10\arcmin$ radius regions of the sensitivity maps. 
We have also calculated $1\sigma$ errors for 
the cumulative distributions following Gehrels (1986).

\begin{figure*}
\centerline{
\includegraphics[scale=0.59]{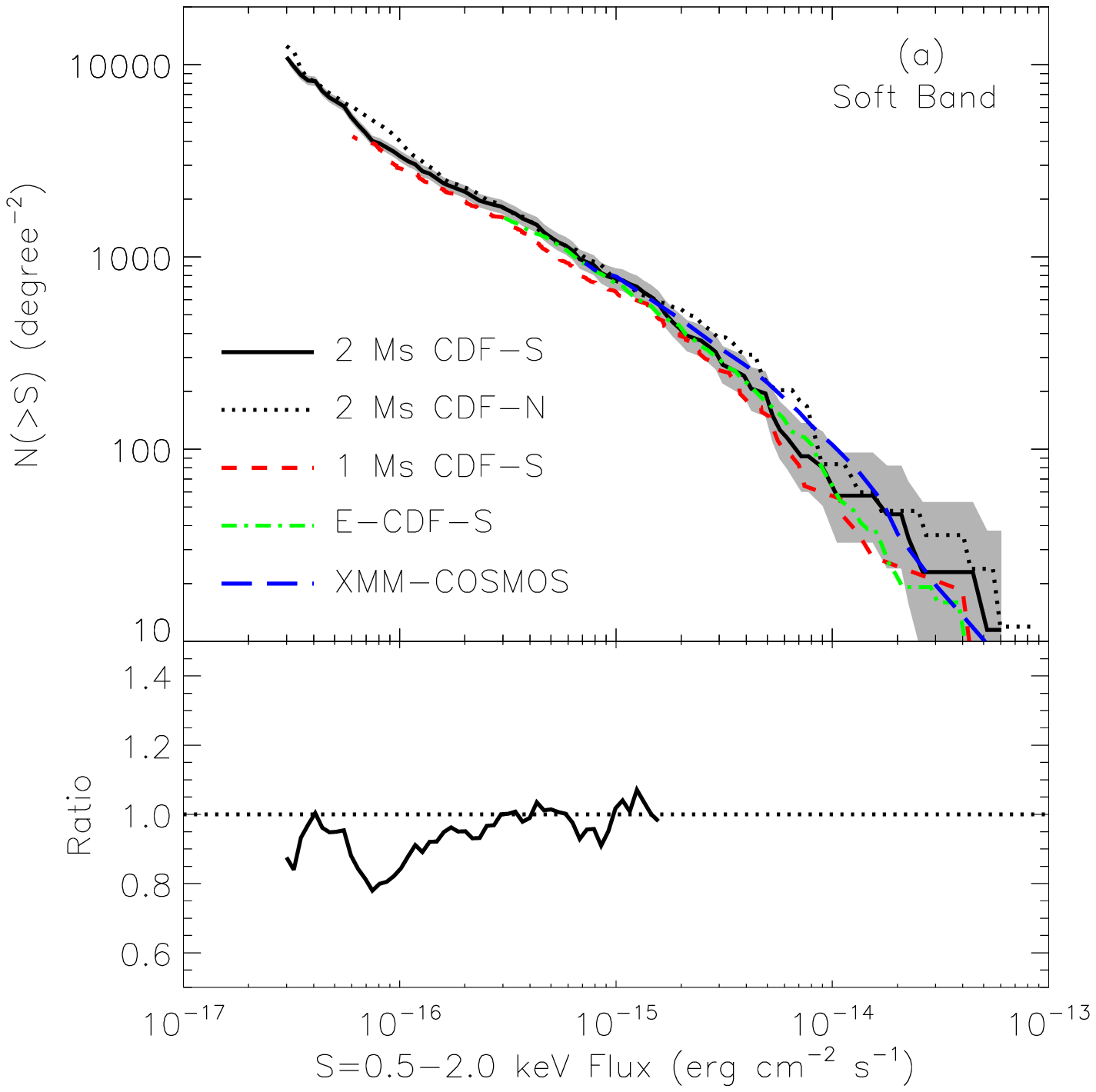}
\includegraphics[scale=0.59]{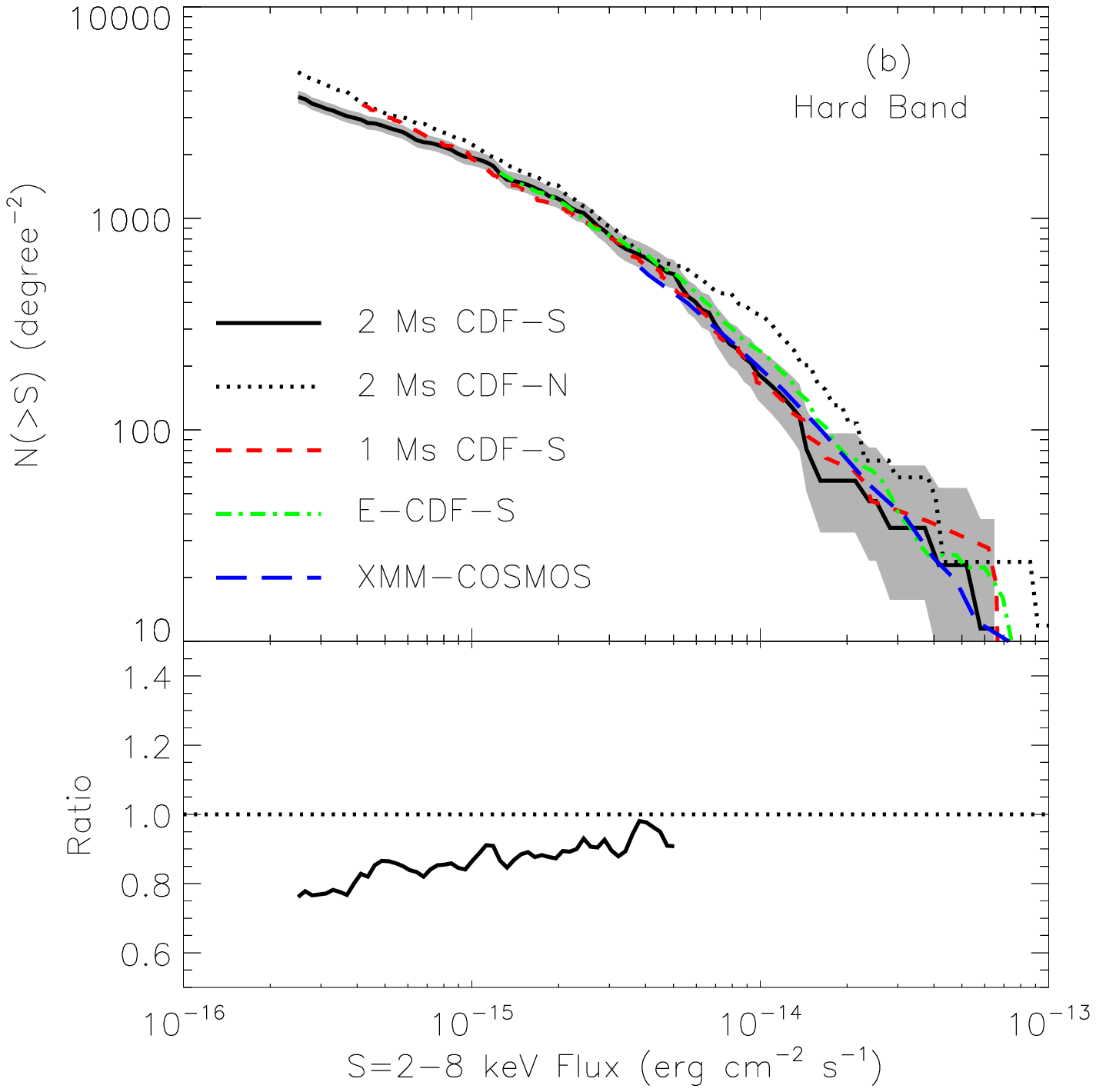}
}
\figcaption{
{\it Top}: Number of sources, $N(>S)$, brighter than a given flux, $S$, for 
the ({\it a}) soft band and ({\it b}) hard band. 
The $\approx$2~Ms \hbox{CDF-S} data 
are plotted as black solid curves with the 1$\sigma$ errors
plotted as gray shaded areas. The cumulative
number counts were computed using 428 \hbox{X-ray} sources in the main 
\chandra\ catalog that are located within $10\arcmin$ of the average 
aim point, and have been corrected for incompleteness and flux bias.
Also shown are the cumulative number-count results for the 
$\approx$2~Ms \hbox{CDF-N} (dotted curves),
the $\approx$1~Ms \hbox{CDF-S} (red dashed curves; \citealt{Rosati2002}),
the $\approx$250~ks \hbox{E-CDF-S} (green dash-dotted curves; L05),
and {\it XMM-Newton} observations in the COSMOS field (blue long-dashed curves; 
\citealt{Cappelluti2007}). The 2--10 keV fluxes in the 
$\approx$1~Ms \hbox{CDF-S} and the COSMOS field were 
converted to 2--8 keV fluxes assuming a photon index $\Gamma=1.4$.
{\it Bottom}: Ratio of the \hbox{CDF-S} to \hbox{CDF-N} cumulative 
number counts for
the ({\it a}) soft band and ({\it b}) hard band. Ratios were calculated 
only for number counts that were derived from a sample of $\ga50$ sources,
since for smaller numbers of sources there are large statistical errors.
This corresponds to soft-band fluxes $\la1.5\times10^{-15}$ \flux\ and
hard-band fluxes $\la5\times10^{-15}$ \flux. 
\label{srccounts}}
\end{figure*}

Figure~\ref{srccounts} displays the cumulative
number counts and the corresponding $1\sigma$ errors for 
the main \chandra\ catalog.
Cumulative number counts for several other surveys have also been shown for
comparison. The derived $\approx$2~Ms \hbox{CDF-S} cumulative
number counts are in general agreement with previous survey results
for the $\approx$1~Ms \hbox{CDF-S}  
\citep{Rosati2002} and the $\approx$250~ks \hbox{E-CDF-S} (L05), at
around the $1\sigma$ confidence level over the entire flux range in
the soft and hard bands. The apparent deviation between the $\approx$2~Ms
and $\approx$1~Ms \hbox{CDF-S} soft-band number counts mainly comes from
the difference in the count-rate--to--flux conversion factor used in these
two surveys.\footnote{An average photon index of $\Gamma=1.4$ was used to 
calculate fluxes in \citet{Rosati2002}, while in this survey, the photon
index was estimated for each source separately and so was the 
count-rate--to--flux conversion factor (see \S3.3.1). We did a test by
calculating the soft-band fluxes using the conversion factor given by 
\citet{Rosati2002}. The derived fluxes are $\sim90\%$ of those 
presented in the main catalog,
and the resulting soft-band number counts are consistent with those for
the $\approx$1~Ms \hbox{CDF-S} to within $1\sigma$.}
The {\it XMM-Newton} observations in the
COSMOS field \citep{Cappelluti2007} provide similar  
number counts, though 
not as deep as the \hbox{CDF-S} observations.

To make a consistent comparison with the $\approx$2~Ms \hbox{CDF-N} 
number counts, we analyzed the \hbox{CDF-N} observations in the same way
as in this paper. A main catalog of 575 \hbox{X-ray} sources 
was constructed. Number counts were calculated using the 496 
X-ray sources located within $10\arcmin$ of the average aim point, and these
have been corrected for incompleteness and flux bias based on simulations.
The \hbox{CDF-N} cumulative number counts  
are presented in Figure~\ref{srccounts} (dotted curves), along with the 
ratios of the \hbox{CDF-S} to \hbox{CDF-N} number counts.
In the soft band, the $\approx$2~Ms \hbox{CDF-S} number counts 
appear to be 
consistent with those for the $\approx$2~Ms \hbox{CDF-N} 
to within $\approx1\sigma$ at fluxes above $\approx2\times10^{-16}$~\flux.
Small differences (up to $\approx3\sigma$) exist at fainter fluxes.
In the hard band, the \hbox{CDF-N} number counts 
deviate above the $1\sigma$ errors of the \hbox{CDF-S} number counts 
at fluxes below $\approx2\times10^{-15}$~\flux; the difference at 
the faintest fluxes is $\approx$25\% ($\approx3\sigma$).
Similar findings of differences between the \hbox{CDF-N} and \hbox{CDF-S}
number counts have been
reported in previous studies \citep[e.g.,][]{Cowie2002,Moretti2003,Bauer2004},
and it appears that this results from small field-to-field variations.
Such field-to-field variations
are generally believed to arise from the large-scale structure underlying the
cosmic \hbox{X-ray} source distribution \citep[e.g.,][]{Gilli2003,Yang2003}.

\section{SUMMARY}
We have presented catalogs and basic analyses of \hbox{X-ray} point sources 
detected in the $\approx2$~Ms \hbox{CDF-S}, which is one of the 
two deepest \chandra\ surveys. 
The key points from this work are the following:

\begin{enumerate}
\item
The entire \hbox{CDF-S} consists of 23 separate observations with
1.911~Ms of combined exposure. The survey covers an area of 435.6 
arcmin$^{2}$.

\item
The main \chandra\ source catalog consists of 462 sources that were
detected using {\sc wavdetect} with a false-positive probability threshold
of $1\times10^{-6}$. These sources were detected in up to three X-ray 
bands: 0.5--8.0~keV, 0.5--2.0~keV,
and 2--8~keV; 135 of these sources are new.

\item
The first supplementary \chandra\ source catalog contains 86 
sources that were generated by
merging the $\approx250$~ks \hbox{E-CDF-S} with the \hbox{CDF-S}, which
provides additional sensitivity in the outer regions of the \hbox{CDF-S}.

\item
The second supplementary \chandra\ source catalog contains 30 sources that were
detected at a lower X-ray significance threshold of $1\times10^{-5}$ 
and that have bright optical counterparts ($R<23.8$).

\item
Source positions for the main 
and supplementary \hbox{CDF-S} plus \hbox{E-CDF-S} \chandra\ catalogs
have been determined using centroid and matched-filter techniques;
the median positional uncertainty is $\approx0\farcs36$.

\item
The basic X-ray and optical properties of the point sources indicate a
variety of source types. More than half of the sources in the
main \chandra\ catalogs
appear to be AGNs. Of the 135 newly detected sources,  
$\approx$55\% appear to be AGNs while $\approx$45\% appear to be starburst
and normal galaxies. 
The majority of the sources in the 
supplementary optically bright catalog are expected to be normal
and starburst galaxies.

\item
The average backgrounds in the
0.5--2.0 and 2--8~keV bands are 0.066 and 0.167~counts Ms$^{-1}$~pixel$^{-1}$,
respectively. Thus these observations are nearly photon limited near the aim
point and could be extended to substantially greater depths with 
further exposure. The background count distributions are very close to 
Poisson distributions. The on-axis flux limits in the 0.5--2.0~keV and
2--8~keV bands are $\approx1.9\times10^{-17}$ \flux\ and
$\approx1.3\times10^{-16}$ \flux, respectively. 

\item
Compared to the other deepest \chandra\ survey, the $\approx2$~Ms \hbox{CDF-N},
the \hbox{CDF-S} has similar effective exposure coverage and sensitivity 
limits. The cumulative number counts of these two fields are consistent
with each other to within $\approx$1~$\sigma$ at fluxes above 
$\approx2\times10^{-16}$~\flux\ in the soft band.
The \hbox{CDF-N} number counts
are up to $\approx25\%$ higher than the  
\hbox{CDF-S} 
number counts at the faintest fluxes in the soft and hard bands, 
indicating small
field-to-field variations.

\end{enumerate}

\section{ACKNOWLEDGEMENTS}
Support for this work was provided by NASA through \chandra\ Award SP8-9003A 
(BL, FEB, WNB) issued by the \chandra\ X-ray Observatory Center, which 
is operated by the Smithsonian Astrophysical Observatory. We also 
acknowledge the financial support of
the Royal Society (DMA and IRS),
the Science and Technology Facilities Council fellowship program (BDL),
NSF grant 06-7634 (DPS), contract ASI--INAF I/023/05/0 and grant 
PRIN-MIUR 2006-02-5203 (AC, RG, and CV). 
We thank H.D. Tananbaum for allocating the time for these observations
and T.~L. Aldcroft, P. Broos and L.~K. Townsley for helpful discussions.



\begin{deluxetable}{lccccccl}
\tabletypesize{\small}
\tablecaption{Journal of {\it Chandra} Deep Field-South Observations}
\tablehead{
\colhead{}                                 &
\colhead{Obs. Start}                                 &
\colhead{Exposure}                             &
\multicolumn{2}{c}{Aim Point$^{\rm b}$}                 &
\colhead{Roll Angle$^{\rm c}$}                                 &
\colhead{Obs.}                                &
\colhead{Pipeline}                             \\
\cline{4-5}
\colhead{Obs. ID}                                 &
\colhead{(UT)}                         &
\colhead{Time$^{\rm a}$ (ks)}               &
\colhead{$\alpha$ (J2000.0)}                &
\colhead{$\delta$ (J2000.0)}                &
\colhead{(deg)}         &
\colhead{Mode$^{\rm d}$}                             &
\colhead{Version$^{\rm e}$}                             
}
\tablewidth{0pt}
\startdata
1431-0\dotfill \ldots \ldots & 1999 Oct 15, 17:38 & \phantom{0}24.6 & 03 32 29.44 & $-$27 48 21.8 & \phantom{0}47.3 & VF & R4CU5UPD11 \\ 
1431-1\dotfill \ldots \ldots & 1999 Nov 23, 02:30 & \phantom{0}93.6 & 03 32 29.44 & $-$27 48 21.8 & 353.9 & \phantom{V}F & R4CU5UPD11 \\       
441\dotfill \ldots \ldots & 2000 May 27, 01:18 & \phantom{0}56.0 & 03 32 26.91 & $-$27 48 19.4 & 166.7 & \phantom{V}F & 7.6.10 \\       
582\dotfill \ldots \ldots & 2000 June 03, 02:38 & 130.6 & 03 32 26.97 & $-$27 48 18.5 & 162.9 & \phantom{V}F & 7.6.10 \\    
2406\dotfill \ldots \ldots & 2000 Dec 10, 23:35 & \phantom{0}29.7 & 03 32 28.33 & $-$27 48 36.5 & 332.2 & \phantom{V}F & 7.6.10 \\    
2405\dotfill \ldots \ldots & 2000 Dec 11, 08:14 & \phantom{0}59.6 & 03 32 28.82 & $-$27 48 43.5 & 331.8 & \phantom{V}F & 7.6.10 \\    
2312\dotfill \ldots \ldots & 2000 Dec 13, 03:28 & 123.7 & 03 32 28.28 & $-$27 48 36.9 & 329.9 & \phantom{V}F & 7.6.10 \\    
1672\dotfill \ldots \ldots & 2000 Dec 16, 05:07 & \phantom{0}95.1 & 03 32 28.73 & $-$27 48 44.5 & 326.9 & \phantom{V}F & 7.6.10 \\    
2409\dotfill \ldots \ldots & 2000 Dec 19, 03:55 & \phantom{0}69.0 & 03 32 28.08 & $-$27 48 38.6 & 319.2 & \phantom{V}F & 7.6.10 \\    
2313\dotfill \ldots \ldots & 2000 Dec 21, 02:08 & 130.4 & 03 32 28.08 & $-$27 48 38.6 & 319.2 & \phantom{V}F & 7.6.10 \\    
2239\dotfill \ldots \ldots & 2000 Dec 23, 17:28 & 130.8 & 03 32 28.08 & $-$27 48 38.6 & 319.2 & \phantom{V}F & 7.6.10 \\    
8591\dotfill \ldots \ldots & 2007 Sep 20, 05:26 & \phantom{0}45.4 & 03 32 28.20 & $-$27 48 06.9 & \phantom{0}72.7 & VF & 7.6.11.1 \\    
9593\dotfill \ldots \ldots & 2007 Sep 22, 20:34 & \phantom{0}46.4 & 03 32 28.20 & $-$27 48 06.9 & \phantom{0}72.7 & VF & 7.6.11.1 \\    
9718\dotfill \ldots \ldots & 2007 Oct 03, 13:56 & \phantom{0}49.4 & 03 32 28.61 & $-$27 48 07.4 & \phantom{0}62.0 & VF & 7.6.11.1 \\    
8593\dotfill \ldots \ldots & 2007 Oct 06, 02:04 & \phantom{0}49.5 & 03 32 28.61 & $-$27 48 07.4 & \phantom{0}62.0 & VF & 7.6.11.1 \\    
8597\dotfill \ldots \ldots & 2007 Oct 17, 07:07 & \phantom{0}59.3 & 03 32 29.25 & $-$27 48 10.4 & \phantom{0}44.2 & VF & 7.6.11.2 \\    
8595\dotfill \ldots \ldots & 2007 Oct 19, 14:16 & 115.4 & 03 32 29.35 & $-$27 48 11.2 & \phantom{0}41.2 & VF & 7.6.11.2 \\    
8592\dotfill \ldots \ldots & 2007 Oct 22, 12:14 & \phantom{0}86.6 & 03 32 29.62 & $-$27 48 13.8 & \phantom{0}32.4 & VF & 7.6.11.2 \\    
8596\dotfill \ldots \ldots & 2007 Oct 24, 13:20 & 115.1 & 03 32 29.62 & $-$27 48 13.8 & \phantom{0}32.4 & VF & 7.6.11.2 \\    
9575\dotfill \ldots \ldots & 2007 Oct 27, 05:43 & 108.7 & 03 32 29.62 & $-$27 48 13.8 & \phantom{0}32.4 & VF & 7.6.11.2 \\    
9578\dotfill \ldots \ldots & 2007 Oct 30, 22:35 & \phantom{0}38.6 & 03 32 29.84 & $-$27 48 16.7 & \phantom{0}24.2 & VF & 7.6.11.2 \\    
8594\dotfill \ldots \ldots & 2007 Nov 01, 11:51 & 141.4 & 03 32 29.84 & $-$27 48 16.7 & \phantom{0}24.2 & VF & 7.6.11.2 \\    
9596\dotfill \ldots \ldots & 2007 Nov 04, 04:11 & 111.9 & 03 32 29.95 & $-$27 48 18.5 & \phantom{0}19.8 & VF & 7.6.11.2 \\    

\enddata

\tablecomments{
The focal-plane temperature was $-110\degr$C during the first two 
observations and $-120\degr$C during the others. Units of right
ascension are hours, minutes, and seconds, and units of declination are 
degrees, arcminutes, and arcseconds.
}
\par \tablenotetext{a}{All observations were continuous. The data were filtered
on good-time intervals, and one mild flare was removed in observation 1431-0.
The short time
intervals with bad satellite aspect are negligible and have not been removed.
The total exposure time for the 23 observations is 1.911~Ms.}
\tablenotetext{b}{The average aim point, weighted by exposure time, is
$\alpha_{\rm J2000.0}=03^{\rm h}32^{\rm m}28\fs80$, 
$\delta_{\rm J2000.0}=-27\degr48\arcmin23\farcs0$.}

\tablenotetext{c}{Roll angle describes the orientation of the \chandra\ 
instruments
on the sky. The angle is between 0--360$^{\circ}$, and it increases to the west
of north (opposite to the sense of traditional position angle).}
\tablenotetext{d}{The observing mode: F=Faint mode and VF=Very Faint mode.}
\tablenotetext{e}{The version of the CXC pipeline software used for basic 
processing of the data.}
\label{tbl-obs}
\end{deluxetable}

\begin{deluxetable}{lllcccccccc}
\tabletypesize{\scriptsize}
\tablewidth{0pt}
\tablecaption{Main {\it Chandra} Catalog}

\tablehead{
\colhead{} &
\multicolumn{2}{c}{X-ray Coordinates} &
\colhead{}                   &
\colhead{}                   &
\multicolumn{6}{c}{Counts}      \\
\\ \cline{2-3} \cline{6-11} \\
\colhead{No.}                    &
\colhead{$\alpha_{2000}$}       &
\colhead{$\delta_{2000}$}       &
\colhead{Pos Err}       &
\colhead{Off-Axis}       &
\colhead{FB}          &
\colhead{FB Upp Err}          &
\colhead{FB Low Err}          &
\colhead{SB}          &
\colhead{SB Upp Err}          &
\colhead{SB Low Err}          \\
\colhead{(1)}         &
\colhead{(2)}         &
\colhead{(3)}         &
\colhead{(4)}         &
\colhead{(5)}         &
\colhead{(6)}         &
\colhead{(7)}         &
\colhead{(8)}         &
\colhead{(9)}        &
\colhead{(10)}        &
\colhead{(11)}
}

\startdata
       1 \dotfill \ldots   &03 31 34.19 &$-$27 50 04.2& 1.6&12.19&     26.1&    11.1&    11.8&     13.5&    $-$1.0&    $-$1.0\\
       2 \dotfill \ldots   &03 31 35.79 &$-$27 51 34.7& 1.9& 12.14&    14.9&    $-$1.0&    $-$1.0&    12.1&    \phantom{0}7.3&    \phantom{0}7.2\\
       3 \dotfill \ldots   &03 31 40.15 &$-$27 47 46.3& 1.3&10.77&     33.6&    11.8&    11.8&     25.8&     \phantom{0}8.4&     \phantom{0}8.1\\
       4 \dotfill \ldots   &03 31 40.93 &$-$27 46 21.8& 1.1& 10.77&     61.2&    14.0&    14.0&     16.0&    $-$1.0&     $-$1.0\\
       5 \dotfill \ldots   &03 31 44.23 &$-$27 49 25.5& 1.0& \phantom{0}9.91&     79.5&    19.4&    19.4&     37.8&     12.5&     12.5\\
\enddata

\tablecomments{
Units of right
ascension are hours, minutes, and seconds, and units of declination are
degrees, arcminutes, and arcseconds.
Table~2 is presented in its entirety in the electronic edition. An 
abbreviated version of the table is shown here for guidance as 
to its form and content. The full table contains 49 columns of 
information on the 462 \hbox{X-ray} sources.}
\label{tbl-mcat}

\end{deluxetable}

\begin{deluxetable}{lccccc}

\tabletypesize{\small}
\tablewidth{0pt}
\tablecaption{Summary of {\it Chandra} Source Detections \label{tbldet}}

\tablehead{
\colhead{} &
\colhead{Number of} &
\multicolumn{4}{c}{Detected Counts Per Source} \\
\cline{3-6}   
\colhead{Band (keV)} &
\colhead{Sources} &
\colhead{Maximum} &
\colhead{Minimum} &
\colhead{Median} &
\colhead{Mean}
}

\startdata
Full (0.5--8.0)   & 403 & 21579.7 & 11.4 & 101.0 & 410.6 \\
Soft (0.5--2.0)  & 392 & 15929.7 & \phantom{0}4.7 & \phantom{0}53.0 & 269.9  \\
Hard (2--8)  & 265 & \phantom{0}5664.3 & \phantom{0}7.7 & \phantom{0}88.6 & 216.9  \\
\enddata
\end{deluxetable}

\begin{deluxetable}{lccc}

\tabletypesize{\small}
\tablewidth{0pt}

\tablecaption{Sources Detected in One Band but not Another \label{tblundet}}

\tablehead{
\colhead{Detection Band} &
\multicolumn{3}{c}{Nondetection Energy Band} \\
\cline{2-4}
\colhead{(keV)} &
\colhead{Full} &
\colhead{Soft} &
\colhead{Hard} 
}

\startdata
Full (0.5--8.0)  & \ldots & 67 & 141 \\
Soft (0.5--2.0)  & ~~~~56~~~~ & \ldots & 166 \\
Hard (2--8)   & ~~~~\phantom{0}3~~~~ & 39  & \ldots \\
\enddata
\tablecomments{For example, there were 67 sources detected in the full band
that were not detected in the soft band.}
\end{deluxetable}

\begin{deluxetable}{lllcccccccc}
\tabletypesize{\scriptsize}
\tablewidth{0pt}
\tablecaption{Supplementary \hbox{CDF-S} plus \hbox{E-CDF-S} \chandra\ Catalog}

\tablehead{
\colhead{} &
\multicolumn{2}{c}{X-ray Coordinates} &
\colhead{}                   &
\colhead{}                   &
\multicolumn{6}{c}{Counts}      \\
\\ \cline{2-3} \cline{6-11} \\
\colhead{No.}                    &
\colhead{$\alpha_{2000}$}       &
\colhead{$\delta_{2000}$}       &
\colhead{Pos Err}       &
\colhead{Off-Axis}       &
\colhead{FB}          &
\colhead{FB Upp Err}          &
\colhead{FB Low Err}          &
\colhead{SB}          &
\colhead{SB Upp Err}          &
\colhead{SB Low Err}          \\
\colhead{(1)}         &
\colhead{(2)}         &
\colhead{(3)}         &
\colhead{(4)}         &
\colhead{(5)}         &
\colhead{(6)}         &
\colhead{(7)}         &
\colhead{(8)}         &
\colhead{(9)}        &
\colhead{(10)}        &
\colhead{(11)}
}

\startdata
       1 \dotfill \ldots   &03 31 40.98 &$-$27 44 34.8& 1.0&11.24&     118.4&    12.8&    11.6&     56.1&    \phantom{0}8.8&    \phantom{0}7.7\\
       2 \dotfill \ldots   &03 31 42.76 &$-$27 53 40.7& 1.6& 11.47&    \phantom{0}17.4&    \phantom{0}5.9&    \phantom{0}4.7&    \phantom{0}7.5&    \phantom{0}4.2&    \phantom{0}3.0\\
       3 \dotfill \ldots   &03 31 43.21 &$-$27 54 05.1& 0.9& 11.58&     152.3&    14.2&    13.1&     49.7&     \phantom{0}8.4&     \phantom{0}7.3\\
       4 \dotfill \ldots   &03 31 44.64 &$-$27 45 19.1& 1.2& 10.23&     \phantom{0}39.9&   \phantom{0}8.1&    \phantom{0}6.9&     \phantom{0}7.5&    $-$1.0&     $-$1.0\\
       5 \dotfill \ldots   &03 31 48.14 &$-$27 52 32.1& 1.6& \phantom{0}9.90&     \phantom{0}10.8&   $-$1.0&    $-$1.0&     \phantom{0}8.1&    \phantom{0}4.4&     \phantom{0}3.2\\
\enddata
\tablecomments{
Units of right
ascension are hours, minutes, and seconds, and units of declination are
degrees, arcminutes, and arcseconds.
Table~5 is presented in its entirety in the electronic edition. An
abbreviated version of the table is shown here for guidance as
to its form and content. The full table contains 52 columns of
information on the 86 \hbox{X-ray} sources.}
\label{tbl-sp1}

\end{deluxetable}

\begin{deluxetable}{lllcccccccc}
\tabletypesize{\scriptsize}
\tablewidth{0pt}
\tablecaption{Supplementary Optically Bright \chandra\ Catalog}

\tablehead{
\colhead{} &
\multicolumn{2}{c}{X-ray Coordinates} &
\colhead{}                   &
\colhead{}                   &
\multicolumn{6}{c}{Counts}      \\
\\ \cline{2-3} \cline{6-11} \\
\colhead{No.}                    &
\colhead{$\alpha_{2000}$}       &
\colhead{$\delta_{2000}$}       &
\colhead{Pos Err}       &
\colhead{Off-Axis}       &
\colhead{FB}          &
\colhead{FB Upp Err}          &
\colhead{FB Low Err}          &
\colhead{SB}          &
\colhead{SB Upp Err}          &
\colhead{SB Low Err}          \\
\colhead{(1)}         &
\colhead{(2)}         &
\colhead{(3)}         &
\colhead{(4)}         &
\colhead{(5)}         &
\colhead{(6)}         &
\colhead{(7)}         &
\colhead{(8)}         &
\colhead{(9)}        &
\colhead{(10)}        &
\colhead{(11)}
}

\startdata
       1 \dotfill \ldots   &03 31 50.82 &$-$27 47 03.8& 1.2&8.50&     47.2&    $-$1.0&    $-$1.0&     22.1&    5.8&    4.7\\
       2 \dotfill \ldots   &03 31 52.03 &$-$27 50 37.6& 1.2& 8.43&    40.9&    $-$1.0&    $-$1.0&    20.8&     5.6&    4.5\\
       3 \dotfill \ldots   &03 31 57.23 &$-$27 45 36.9& 1.2&7.51&     41.6&    $-$1.0&    $-$1.0&     22.7&     5.8&     4.7\\
       4 \dotfill \ldots   &03 32 00.32 &$-$27 46 11.4& 1.2& 6.67&     35.9&   $-$1.0&    $-$1.0&     18.7&    5.4&     4.3\\
       5 \dotfill \ldots   &03 32 06.59 &$-$27 50 37.3& 1.2& 5.39&     24.3&   $-$1.0&    $-$1.0&     12.0&    4.6&     3.4\\
\enddata
\tablecomments{
Units of right
ascension are hours, minutes, and seconds, and units of declination are
degrees, arcminutes, and arcseconds.
Table~6 is presented in its entirety in the electronic edition. An
abbreviated version of the table is shown here for guidance as
to its form and content. The full table contains 38 columns of
information on the 30 \hbox{X-ray} sources.}
\label{tbl-sp2}

\end{deluxetable}

\begin{deluxetable}{lcccc}

\tabletypesize{\footnotesize}
\tablecaption{Background Parameters}
\tablehead{
\colhead{}                                 &
\multicolumn{2}{c}{Mean Background}                 &
\colhead{Total Background$^{\rm c}$}                                 &
\colhead{Count Ratio$^{\rm d}$}                                \\
\cline{2-3}
\colhead{Band (keV)}                                 &
\colhead{(counts pixel$^{-1}$)$^{\rm a}$}                         &
\colhead{(counts Ms$^{-1}$ pixel$^{-1}$)$^{\rm b}$}               &
\colhead{(10$^5$ counts)}                &
\colhead{(background/source)}                
}
\tablewidth{0pt}
\startdata
Full (0.5--8.0)  & 0.248 & 0.242  & 16.1 & \phantom{0}9.7  \\
Soft (0.5--2.0)   & 0.067 & 0.066  & \phantom{0}4.3 & \phantom{0}4.1  \\
Hard (2--8)   & 0.179 & 0.167  & 11.6 & 20.2  \\
\enddata
\label{tbl-bkg}
\par \tablenotetext{a}{The mean numbers of background counts per pixel. These are 
measured from the background images described in $\S$4.}
\tablenotetext{b}{The mean numbers of counts per pixel 
divided by the mean effective exposure. 
These are measured from the exposure maps and 
background images described in $\S$4.}
\tablenotetext{c}{Total number of background counts.}
\tablenotetext{d}{Ratio of the total number of background 
counts to the total number of source counts.}

\end{deluxetable}

\end{document}